\documentclass[conference]{IEEEtran}
\IEEEoverridecommandlockouts
\usepackage{cite}
\usepackage{amsmath,amssymb,amsfonts}
\usepackage{algorithmic}
\usepackage{graphicx}
\usepackage{textcomp}
\usepackage{xcolor}
\def\BibTeX{{\rm B\kern-.05em{\sc i\kern-.025em b}\kern-.08em
    T\kern-.1667em\lower.7ex\hbox{E}\kern-.125emX}}

\usepackage{tikz}
\usepackage{graphicx}
\usepackage{amsmath}
\usepackage{amsthm}
\usepackage{amsfonts,amssymb}
\usepackage{algorithm}
\usepackage{algorithmic}
\usepackage{framed}
\usepackage{mathrsfs}
\usepackage{enumerate}
\usepackage{xcolor}
\usepackage{lipsum} 
\usepackage{tabulary}
\usepackage{booktabs}
\usepackage{array}
\usepackage{subcaption}
\usepackage{filecontents}
\usepackage{setspace}
\usepackage[normalem]{ulem}

\usepackage{bm}
\usepackage{multirow}

\newtheorem{definition}{Definition}
\newtheorem{lemma}{Lemma}
\newtheorem{theorem}{Theorem}
\newtheorem{corollary}{Corollary}
\newtheorem{assumption}{Assumption}
\newtheorem{convention}{Convention}

\usepackage[draft]{todonotes}   

\newcommand{\structure}[1]{{\color{red}{}}}
\newcommand{\comment}[1]
{{\color{black}#1}} 

\newcommand{\SWITCH}[1]{\STATE \textbf{switch} (#1) \begin{ALC@g}}
\newcommand{\ENDSWITCH}{\end{ALC@g}\STATE \textbf{end switch}}
\newcommand{\CASE}[1]{\STATE \textbf{case} #1\textbf{:} \begin{ALC@g}}
\newcommand{\ENDCASE}{\end{ALC@g}}

\newcommand{\DEFAULT}{\STATE \textbf{default:} \begin{ALC@g}}
\newcommand{\ENDDEFAULT}{\end{ALC@g}}
\newcommand{\DEFAULTLINE}[1]{\STATE \textbf{default:} }

\newcommand{\hide}[1]{}
\newcommand{\hidemass}[1]{}

\begin{document}

\title{TPMDP: Threshold Personalized Multi-party Differential Privacy via Optimal Gaussian Mechanism
}

\author{
\IEEEauthorblockN{
Jiandong~Liu\textsuperscript{1},
Lan~Zhang\textsuperscript{1},
Chaojie~Lv\textsuperscript{1},
Ting~Yu\textsuperscript{2},
Nikolaos~M.~Freris\textsuperscript{1},
and Xiang-Yang~Li\textsuperscript{1}
}
\IEEEauthorblockA{\textsuperscript{1}{University of Science and Technology of China, Hefei, China} \\
\textsuperscript{2}{Hamad Bin Khalifa University, Qatar} \\
jdliu@mail.ustc.edu.cn, zhanglan@ustc.edu.cn, chjlv@mail.ustc.edu.cn, tyu@hbku.edu.qa,\\
nfr@ustc.edu.cn, xiangyangli@ustc.edu.cn}
}

\maketitle

\begin{abstract}

In modern distributed computing applications, such as federated learning and AIoT systems, protecting privacy is crucial to prevent adversarial parties from colluding to steal others' private information. However, guaranteeing the utility of computation outcomes while protecting all parties' data privacy can be challenging, particularly when the parties' privacy requirements are highly heterogeneous. In this paper, we propose a novel privacy framework for multi-party computation called Threshold Personalized Multi-party Differential Privacy (TPMDP), which addresses a limited number of semi-honest colluding adversaries. Our framework enables each party to have a personalized privacy budget. We design a multi-party Gaussian mechanism that is easy to implement and satisfies TPMDP, wherein each party perturbs the computation outcome in a secure multi-party computation protocol using Gaussian noise. To optimize the utility of the mechanism, we cast the utility loss minimization problem into a linear programming (LP) problem. We exploit the specific structure of this LP problem to compute the optimal solution after $\mathcal{O}(n)$ computations, where $n$ is the number of parties, while a generic solver may require exponentially many computations. Extensive experiments demonstrate the benefits of our approach in terms of low utility loss and high efficiency compared to existing private mechanisms that do not consider personalized privacy requirements or collusion thresholds.
\end{abstract}

\begin{IEEEkeywords}
Differential privacy, secure multi-party computation, personalized privacy, distributed computing.
\end{IEEEkeywords}

\section{Introduction}\label{sec:intro}
\structure{
Define private collaborative computing.
Example of private collaborative computing.
Utility objective in private collaborative computing.
Define our problem.
}


Collaborative computing, which involves multiple parties, has garnered significant attention in
distributed systems like IoT and ad-hoc networks.
This approach enables parties to obtain results with higher utility or accuracy based on more abundant datasets in tasks such as distributed computing~\cite{WoolseyWCJ23},
edge computing~\cite{RubleinMT0P21}, and federated learning~\cite{0004SM20}.
However, protecting privacy of the involved parties who contribute data in computing is crucial. For instance, in collaborative recommendations~\cite{LinLDQPR23}, multiple parties, such as video websites, banks, and online shops, contribute data to train an AI model to predict
clients' preferences. Each party holds a personalized privacy requirement, and the privacy requirement of banks, for example, can be more stringent than
that of
video websites.


Several solutions for (personalized) privacy-preserving collaborative computing
have been proposed in the current literature~\cite{nissim2011distributed,acar2017achieving,Murakami019,cheu2019distributed} based on differential privacy (DP)~\cite{dwork2014algorithmic}. These solutions aim to ensure the privacy of all parties by perturbing the outputs or inputs in collaborative computing. However, these methods can result in a prohibitively high noise level, which may significantly compromise the utility of the output. In this work, we propose a new personalized privacy framework for collaborative computing that offers higher utility.

\structure{
Introduce the general theory.
Give insights of designing
algorithms.
}


To solve the problem of personalized private collaborative computing, we propose a new concept called threshold personalized multi-party differential privacy (TPMDP). TPMDP \comment{draws on} insights from both threshold secure multi-party computation (t-MPC)~\cite{goldreich2007foundations} and personalized differential privacy (PDP)~\cite{jorgensen2015conservative}. t-MPC provides a collaborative computing framework that allows multiple parties to compute accurate results while ensuring that no collusion of $t$ or fewer parties learns more than the outcome, where $t$ is a predefined threshold value. In real-world applications, parties may be geographically isolated or have conflicting interests, which limits their ability to collude arbitrarily (e.g., distributed learning in Internet-of-Things (IoT) systems~\cite{Jiang0Z020}). TPMDP ensures that all parties' personalized DP requirements are met for all collusions of size up to a given threshold $t$.

\structure{
Introduce algorithms. Explain why and how we address the
problems.
}

\IEEEpubidadjcol


The benefits of the TPMDP framework are demonstrated through the design of a highly effective and easy-to-implement multi-party Gaussian mechanism that satisfies TPMDP. In this mechanism, each participating party samples a Gaussian noise that adheres to all parties' privacy requirements. Following this, all parties input their respective data and noise to a t-MPC protocol, which computes the query function perturbed by the \comment{noises} contributed by all parties.


The utility of the multi-party Gaussian mechanism is optimized using an efficient parameter-choosing algorithm. We transform the problem of minimizing utility loss measured by the noise variance into a linear programming (LP) problem, which is characterized by numerous constraints that cannot be efficiently solved using a generic LP solver. To address this challenge, we leverage the structure of the LP problem and propose a method that yields the optimal solution with linear complexity $\mathcal{O}(n)$, where $n$ represents the number of parties.

\structure{Introduce the theoretical advantages of our algorithms.}


Our multi-party Gaussian mechanism offers notable theoretical advantages. In particular, compared to the threshold multi-party DP (TMDP) mechanism~\cite{nissim2011distributed}, where each party adds noise with uniform variance determined by their most stringent privacy requirement, our mechanism achieves superior utility, especially in scenarios where parties have heterogeneous privacy requirements and the collusion threshold is large. On the other hand, compared with the personalized local DP (PLDP) mechanism~\cite{Murakami019} without considering collusion thresholds, the advantage of the multi-party Gaussian mechanism is prominent when the threshold $t \leq p n$ for a constant $p < 1$ as $n$ increases.

For the LP problem of maximizing the multi-party Gaussian mechanism's utility,
we compare the efficiency of our solver with that of the state-of-the-art generic LP solver~\cite{lee2019solving}. We find that the \comment{space and time} complexity of the generic solver can be exponential in $n$. In contrast, our solver computes the optimal variance assignment with
$\mathcal{O} (n)$ storage and running~time.

\structure{Introduce the experiments.}


We conduct experiments on synthetic and real-world datasets, which demonstrate that our TPMDP multi-party Gaussian mechanism outperforms both the TMDP and PLDP mechanisms in terms of utility.
Moreover, we compare our mechanism with centralized (personalized) DP mechanisms where a trusted third party gathers data from all parties, computes the query function with additive noise, and then releases the results.
It shows that our mechanism achieves utility comparable to centralized mechanisms when
the honest parties comprise the majority (i.e., $t < 0.5n$).

Furthermore, we highlight the scalability advantages of our exact algorithm for solving the LP problem. Our method outperforms generic LP solvers and can efficiently handle many parties, making it a practical and scalable solution for multi-party privacy mechanisms.

\section{Preliminaries}\label{sec:preliminary}

In this section, we introduce pertinent concepts from secure multi-party computation (MPC) and differential privacy (DP).
\subsection{Privacy of MPC}

\hide{
\begin{table}[h]
\centering
{
        \begin{tabular}{>{\centering\arraybackslash} p{1.75cm}   m{5.85cm} }
        \toprule
        Symbol & Meaning\\
        \midrule
        $U$        & Set of parties in an MPC protocol\\
        $A$ & Adversarial group \\
        $\mathcal{A}$, $\mathcal{A}_t(\mathcal{A}_{\tau})$ & Adversary structure, $t$($\tau$)-threshold adversary structure \\
        $\mathcal{V}_{A}^{\Pi}({\boldsymbol{x}})$ & Viewed values of parties in $A$ during an execution of protocol $\Pi$ with input ${\boldsymbol{x}}$ \\
        $\overset{c}{\equiv}$ / $\overset{s}{\equiv}$ & Computationally / statistically indistinguishable \\
        \bottomrule
        \end{tabular}
        \caption{Notation in MPC.}
        \label{tab:notation}
        }
\end{table}
}

This paper focuses on MPC protocols with $n$ semi-honest parties. Semi-honest parties follow the protocol but attempt to infer others' private data through the messages they obtain during the protocol execution, denoted as their views.

Given an $n$-ary deterministic function $f:\ (\{0,1\}^{*})^n\rightarrow (\{0,1\}^{*})^n$, we denote $f_A({\boldsymbol{x}})$ as
$\{f_i (\boldsymbol{x})\}_{i\in A}$,
where ${\boldsymbol{x}}\triangleq (x_1,\hdots,x_n)$, $f_i({\boldsymbol{x}})$ denotes the $i$-th element of $f({\boldsymbol{x}})$, and $A \subset [n]$ is an adversarial group.
Here, $\{0, 1\}^*$ denotes the space of binary strings with arbitrary length.
The view of party $j\in [n]$ on the MPC protocol $\pi$ with input ${\boldsymbol{x}}$, denoted by $\mathcal{V}_{j}^{\pi}({\boldsymbol{x}})$, is the vector consisting of all the messages that party $j$ inputs and receives during the execution of $\pi$. The view of the adversarial group $A$ is denoted as
$\mathcal{V}_A^\pi({\boldsymbol{x}}) \triangleq (A, \{\mathcal{V}_{i}^{\pi} ({\boldsymbol{x}})\}_{i\in A})$.

A protocol $\pi$ privately computes $f$ in the presence of any adversarial group $A$ if $A$'s view $\mathcal{V}_A^\pi({\boldsymbol{x}})$ can be essentially computed from the inputs and outputs available to $A$, such that computationally bounded or unbounded adversaries cannot distinguish it. The formal definitions of computational and statistical indistinguishability (denoted by $\overset{c}{\equiv}$ and $\overset{s}{\equiv}$, which provide privacy against computationally bounded and unbounded adversaries, respectively) can be found in~\cite{goldreich2007foundations}. We define the adversary structure $\mathcal{A}$ as the set of all considered adversarial groups $A$. We say that protocol $\pi$ privately computes $f$ if it privately computes $f$ in the presence of $A$ for all $A\in \mathcal{A}$.

\begin{definition}[Privacy of $n$-party MPC protocols for deterministic functions in semi-honest settings~\cite{goldreich2007foundations}]\label{def:c_MPC}
    We say that an MPC protocol \textbf{$\pi$ computationally $t$-privately computes} the deterministic function $f$ if there exists a probabilistic polynomial-time algorithm $\mathcal{S}$ such that for $A \in \mathcal{A}_t\triangleq \{A \subset [n] \mid |A| \leq t\}$,
    %
    \begin{equation}\label{eq:c_deter_MPC}
        \begin{aligned}
            \{\mathcal{S}(A,\{x_i\}_{i\in A},  f_A({\boldsymbol{x}}))\}_{{\boldsymbol{x}}\in (\{0,1\}^{*})^n} \overset{c}\equiv
            \{\mathcal{V}_A^\pi ({\boldsymbol{x}})\}_{{\boldsymbol{x}}\in (\{0,1\}^{*})^n}.
        \end{aligned}
    \end{equation}
    Similarly, we say that $\pi$ \textbf{statistically $t$-privately} computes $f$, if for every $A\in \mathcal{A}_t$, the left-hand and right-hand sides in Eq.~\eqref{eq:c_deter_MPC} are statistically
    indistinguishable.
\end{definition}
%
%


In computationally private MPC, privacy is maintained even in unlimited collusions ($t = n-1$). On the other hand, in statistically private MPC, privacy is only guaranteed when the honest parties constitute a strict majority ($t < n/2$)
\cite{goldreich2007foundations}.

\hide{
\begin{table}[t]
\centering
{
        \begin{tabular}{>{\centering\arraybackslash} p{1.1 cm}  m{6.5cm} }
        \toprule
        Symbol & Meaning \\
        \midrule
        $(\epsilon, \delta)$ & Privacy budget in DP \\
        $\Delta_2 f$ & $l_2$-sensitivity of function $f$ \\
        $\Gamma$ & The tuple $(\epsilon, \delta, \Delta_2 f)$ \\
        $\sigma_{\Gamma}$ & The minimum $\sigma$ of $(\epsilon, \delta)$-DP Gaussian mechanisms with sensitivity $\Delta_2 f$\\
        $\simeq$ & Neighboring relation, e.g. $x$ and $y$ satisfy $x\simeq y$ if $x$ and $y$ differ in one single element\\
        \bottomrule
        \end{tabular}
        \caption{Notation in DP.}
        \label{tab:notation_DP}
        }
\end{table} 
}

\vspace{-0.1in}
\subsection{DP and Gaussian Mechanism}

In DP, our goal is to design a mechanism $M:\mathcal{X}\rightarrow\mathcal{Y}$ that ensures the privacy of each record in an arbitrary collection ${\boldsymbol{x}}\in \mathcal{X}$. Here, $\mathcal{X}$ denotes the set of all possible input collections. We use the notation ${\boldsymbol{x}}\simeq\boldsymbol{y}$ to represent neighboring input collections ${\boldsymbol{x}},{\boldsymbol{y}}\in \mathcal{X}$, i.e., inputs that are either identical or differ by at most one element.

\begin{definition}[Differential Privacy~\cite{dwork2014algorithmic}]\label{def:DP}
    For $\epsilon\geq 0$, $\delta \in [0,1]$, a randomized algorithm $M$ is $(\epsilon, \delta)$-differentially private if for all $S \subset Range(M)$ and for all ${\boldsymbol{x}}, {\boldsymbol{y}} \in \mathcal{X}$ such that ${\boldsymbol{x}}\simeq {\boldsymbol{y}}$: 
    \begin{equation}\label{eq:DP}
        \Pr[M({\boldsymbol{x}}) \in S] \leq \exp(\epsilon) \Pr[M({\boldsymbol{y}}) \in S] + \delta.
    \end{equation}
\end{definition}

Output perturbation is an important class of DP mechanisms. Let
$f$
be a query function. An output perturbation mechanism adds a noise vector $X$ to $f({\boldsymbol{x}})$, i.e., $M({\boldsymbol{x}})=f({\boldsymbol{x}})+X$. The Gaussian mechanism is a widely adopted output perturbation mechanism, where $X\sim \mathcal{N}(0,\sigma^2 I)$, with $I$ representing a $d$-dimensional identity matrix, where $d$ is the dimension of $f$'s output.
Balle et al.~\cite{balle2018improving} proposed the analytical Gaussian mechanism that is tight for $(\epsilon,\delta)$-DP by selecting an appropriate $\sigma^2$. For a function $f$ with $l_2$-sensitivity
$\Delta_2 f \triangleq \mathop{\max}_{{\boldsymbol{x}}\simeq\boldsymbol{y}}\left\|f({\boldsymbol{x}})-f({\boldsymbol{y}})\right\|_2$
and given $\epsilon$ and $\delta$, one can compute $\sigma_{(\epsilon,\delta, \Delta_2 f)}^2$ using an efficient numerical algorithm with $\mathcal{O}(1)$ complexity. The Gaussian mechanism is $(\epsilon, \delta)$-differentially private if and only if $\sigma^2 \geq \sigma_{(\epsilon,\delta, \Delta_2 f)}^2$.
We use the notations $\Gamma\triangleq (\epsilon, \delta, \Delta_2 f)$ and $\sigma_{(\epsilon,\delta, \Delta_2 f)}^2 \triangleq \sigma_{\Gamma}^2$.

\hide{
\begin{definition}\label{def:sigma_Gamma}
    Given the tuple $\Gamma\triangleq (\epsilon, \delta, \Delta_2 f)$ ,$\sigma_{\Gamma}$ is the lower bound of $\sigma$ such that the Gaussian mechanism with the additive noise $X\sim \mathcal{N}(0, \sigma^2I)$ is $(\epsilon, \delta)$-differentially private.
\end{definition}
}

\hide{
It holds that {$\sigma_{\Gamma}$ is continuous with respect to $\epsilon$, $\delta$, $\Delta_2 f$. Furthermore, it decreases monotonically with $\epsilon$, $\delta$ and increases monotonically with $\Delta_2 f$}. 
}



\section{Problem Formulation}
\label{sec:prob-formulation}
\begin{figure}[!t]
\centering
\includegraphics[width=0.8\linewidth]{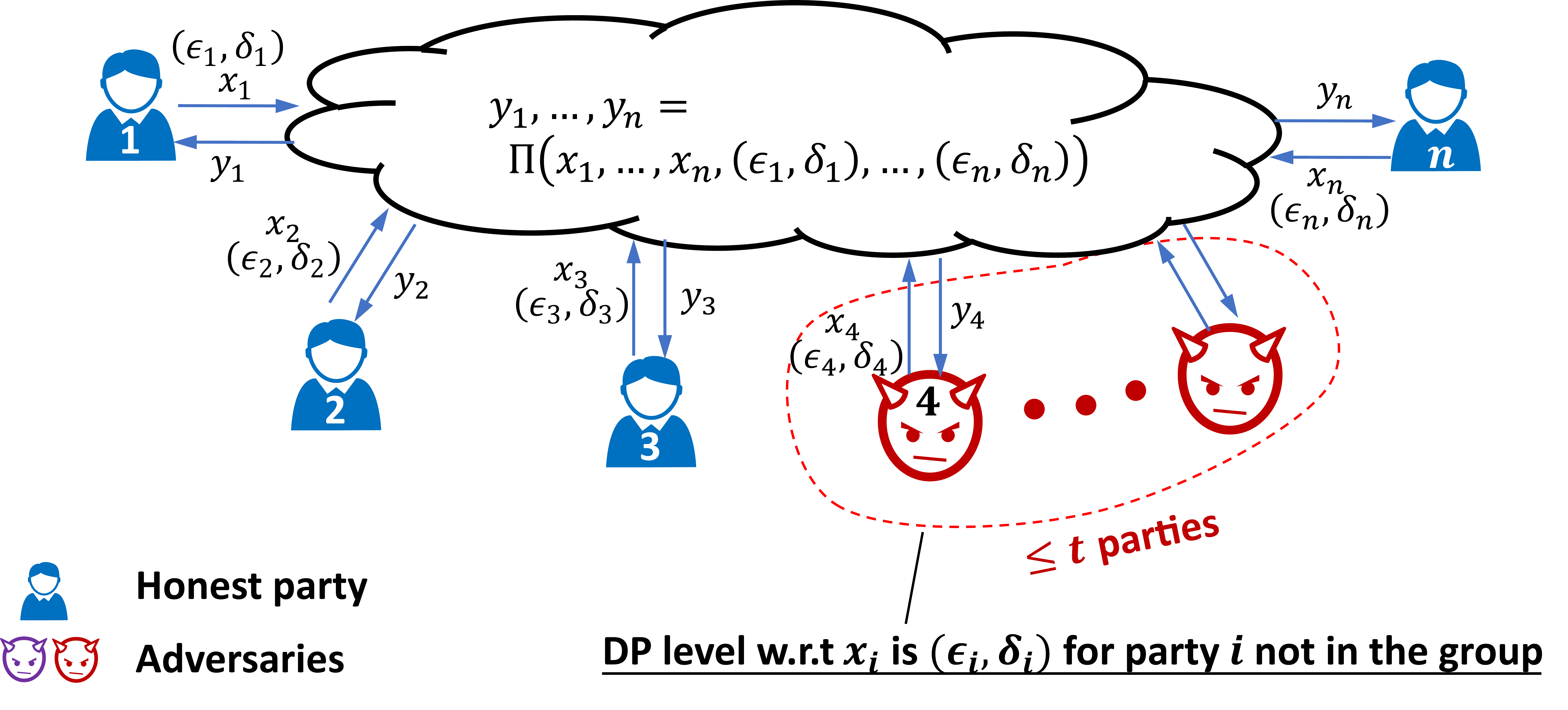}
\caption{
The system model of TPMDP.
The parties are assumed semi-honest, with the size of any adversarial group being limited to $t$. Each party $i$ holds an input $x_i$ and a privacy budget parameterized by $(\epsilon_i, \delta_i)$. All parties execute the protocol $\Pi$ and generate an output $y_i$ for each $i$. The privacy of each party's input is preserved, ensuring that any adversarial group cannot obtain any information about $x_i$ beyond $(\epsilon_i, \delta_i)$-DP, provided that party $i$ is not part of that group.
}\label{fig:sys_setting}
\vspace{-0.1in}
\end{figure}

In this section, we formulate the definition of threshold personalized multi-party differential privacy (TPMDP).

\subsection{Computation and Privacy Threat Model}



As shown in Fig.~\ref{fig:sys_setting}, the TPMDP system model involves a group of parties $[n]$ who seek to privately compute a deterministic function $f(x_1, \dots, x_n)$, where each party $i\in  [n]$ holds an input $x_i \in \{0,1\}^*$. We assume that the parties are semi-honest and that the maximum number of colluding adversaries is $t$.

To quantify the privacy requirements of each party $i$, we use the DP parameters $(\epsilon_i, \delta_i)$ for $i\in [n]$. Specifically, in Definition \ref{def:DP}, we take the function $M$ as an arbitrary inference function and the input $x$ or $y$ as the messages that any adversary can obtain in a TPMDP protocol. We require that Eq.~\eqref{eq:DP} holds for each $i$ to ensure the privacy of party~$i$.
For convenience, we denote $\mathcal{E}\triangleq \{(\epsilon_i, \delta_i)\}_{i\in [n]}$.



\subsection{Definition of TPMDP}


TPMDP provides a metric for the privacy loss of a multi-party computation algorithm in case
at most one party's input varies (denoted by neighboring inputs).
The definition of neighboring inputs is provided in Definition~\ref{def:A_neigh_in}.
We define the colluding group's refined view, as outlined in Definition~\ref{def:rView}, regarding what they can infer from the algorithm run. Finally, we summarize the definition of TPMDP in Definition~\ref{def:tMDP}.

\structure{Theoretical properties and advantages of TPMDP}

\begin{definition}[$j$-neighboring inputs]\label{def:A_neigh_in}
    For $j\in [n]$, two inputs ${\boldsymbol{x}}\triangleq (x_1, \hdots, x_n)$ and ${\boldsymbol{x}}'\triangleq (x_{1}', \hdots, x_n')$ are $j$-neighboring if
    $x_i = x_i'$ for $i\neq j$; we use the notation $x\overset{j}{\simeq} y$.
\end{definition}


\begin{definition}[Refined view]\label{def:rView}
    Let $\mathcal{V}_A^\Pi ({\boldsymbol{x}},{\boldsymbol{X}})$ be the viewed messages of group $A\subset [n]$ in a $n$-party computation protocol $\Pi$ that computes a deterministic function given the input $\boldsymbol{x}$ and collection of randomization terms $\boldsymbol{X}\in supp(\mathcal{P})$.
    Here, $\mathcal{P}$ denotes the distribution of the collection of randomization terms, and $supp(\mathcal{P})$ denotes the support of $\mathcal{P}$.
    A tuple $\mathcal{RV}_A^\Pi ({\boldsymbol{x}},{\boldsymbol{X}})$ is called~a \textbf{computational refined view} for
    $A\subset [n]$ if there exists a probabilistic polynomial-time algorithm $\mathcal{S}$ such that:
    \begin{equation}\label{eq:rView}
        \begin{aligned}
            \{\mathcal{S}(A,\mathcal{RV}_A^\Pi ({\boldsymbol{x}},{\boldsymbol{X}}))  \}_{{\boldsymbol{x}}\in (\{0,1\}^{*})^n,{\boldsymbol{X}}\in supp(\mathcal{P})} \overset{c}\equiv 
            \\ \{\mathcal{V}_A^\Pi ({\boldsymbol{x}},{\boldsymbol{X}})\}_{{\boldsymbol{x}}\in (\{0,1\}^{*})^n,{\boldsymbol{X}}\in supp(\mathcal{P})}.
        \end{aligned}
    \end{equation}
    If the left-hand and right-hand sides in Eq.~\eqref{eq:rView} are \textbf{statistically indistinguishable},
    $\mathcal{RV}_A^\Pi ({\boldsymbol{x}},{\boldsymbol{X}})$ is a \textbf{statistical refined view}.
\end{definition}


\begin{definition}[$(t,n,\mathcal{E})$-TPMDP]\label{def:tMDP}
A protocol $\Pi$ satisfies \textbf{computational $(t,n,\mathcal{E})$-TPMDP} if for each $A\in\mathcal{A}_t = \{A\subset [n] \mid |A| \leq t\}$, there exists a \textbf{computational refined view} $\mathcal{RV}_A^\Pi({\boldsymbol{x}},{\boldsymbol{X}})$ such that for all $j\in \bar{A}$, for all $S \subset \{\mathcal{RV}_A^\Pi({\boldsymbol{x}}, {\boldsymbol{X}}) \}_{{\boldsymbol{x}} \in {(\{0,1\}^*)}^n, {\boldsymbol{X}}\sim \mathcal{P}}$,
and for all $j$-neighboring inputs ${\boldsymbol{x}}$ and ${\boldsymbol{x}}'$:
\begin{equation}\label{eq:tMDP}
\begin{aligned}
    \Pr[\mathcal{RV}_A^\Pi({\boldsymbol{x}},  {\boldsymbol{X}}) \in S] \leq
    e^{\epsilon_j} \Pr[\mathcal{RV}_A^\Pi({\boldsymbol{x}}',{\boldsymbol{X}}) \in S]+\delta_j,
\end{aligned}
\end{equation}
where $\boldsymbol{X}$ denotes the collection of randomization terms in the protocol $\Pi$.
%
When there exists a \textbf{statistical refined view} for each $A\in\mathcal{A}_t$ satisfying Eq.~\eqref{eq:tMDP}, then we say the protocol $\Pi$ satisfies \textbf{statistical $(t,n,\mathcal{E})$-TPMDP.}

\end{definition}

Definition~\ref{def:tMDP} implies that, in a TPMDP mechanism, a set of adversaries $A\in \mathcal{A}_{t}$ cannot distinguish with high certainty between the refined views for neighboring inputs. As the refined view computationally (statistically) covers the information in the running of the mechanism,
we can ensure the privacy of all parties in the computational or statistical sense.

\comment{\textbf{Comparisons with existing works:}}
In comparison to existing definitions of multi-party DP in~\cite{mironov2009computational,nissim2011distributed,Murakami019}, our proposed definition is more comprehensive while remaining consistent with them. Specifically, when degenerating to the non-personalized setting (i.e.,
$(\epsilon_i, \delta_i) \equiv (\epsilon, \delta)$ for some $(\epsilon, \delta)$),
our definition of computational TPMDP is equivalent to the definition of computational multi-party DP (SIM-CDP) as presented in~\cite{mironov2009computational}. Additionally, by leveraging the post-processing property of DP~\cite{dwork2014algorithmic}, we can establish the consistency of our definition with TMDP in~\cite{nissim2011distributed}, which is solely based on the distribution of the view of parties when TPMDP degenerates to the non-personalized setting.
\comment{Regarding the comparison with the definition for PLDP~\cite{Murakami019}, it is a special case for statistical TPMDP with $t = n-1$.}

\hide{
\begin{table}[t]
\centering
{
    \begin{tabular}{>{\centering\arraybackslash} p{1.1cm}  m{6.8cm} }
    \toprule
    Symbol & Meaning \\
    \midrule
    $\mathcal{RV}_{A}^{\Pi}({\boldsymbol{x}})$ & Refined view for $A$ during the execution of a TPMDP mechanism $\Pi$ with input ${\boldsymbol{x}}$, cf. Definition \ref{def:rView} \\
    $\mathcal{E}$ & The tuple of privacy budgets of parties in $U$ \\
    $\perp$ & An information-less symbol \\
    $\Delta_{2,j} F$ & $j$-th partial $l_2$-sensitivity of $F$, cf. Definition \ref{def:p_l2_sensitivity} \\
    ${U^+}$ & Set of active parties receiving outputs unequal to $\perp$ in a multi-party Gaussian mechanism\\
    ${\mathcal{A}_{t}^+}$ & Active $t$-threshold adversary structure consisting of all adversarial groups containing at least one active party\\
    $\boldsymbol{\sigma}$ & The tuple of the standard deviations of noises in a multi-party Gaussian mechanism \\
    $v_{\boldsymbol{\sigma}}$ & The objective function in Eq.~\eqref{eq:Gauss_opt}, i.e., $v_{\boldsymbol{\sigma}} = \Sigma_{i\in U} \sigma_i^2$ \\
    $\overset{j}{\simeq}$ & $j$-neighboring relation, cf. Definition \ref{def:A_neigh_in} \\
    \bottomrule
    \end{tabular}
    \caption{Notation in TPMDP.}
    \label{tab:notation_TPMDP}
    }
\end{table}
}


\section{Multi-party Gaussian Mechanism}
\label{sec:res}

We present an easy-to-implement multi-party Gaussian mechanism that is well-suited for TPMDP with different parameters $(t,n,\mathcal{E})$. 
We choose to use the Gaussian mechanism for several reasons: 1) Gaussian noise is prevalent and can be easily generated in both software and hardware~\cite{abadi2016deep,iyengar2019towards}; 2) Gaussian noise possesses the property that the sum of independent Gaussian noises is also Gaussian; and 3) Gaussian noise simplifies the analysis of mechanism utility, as the problem of minimizing utility loss measured by the overall noise variance can be cast into a linear programming (LP) problem. However, the LP problem for finding the optimal variances for the multi-party Gaussian mechanism includes an exponential number of constraints,
making it computationally expensive to solve using the state-of-the-art generic LP solver~\cite{lee2019solving},
\comment{which requires storage and running time exponential in $n$.}
To address this challenge, we exploit the structure of the LP problem and develop an efficient solver that can solve the problem with $\mathcal{O}(n)$ storage and running time.
The utility of our mechanism surpasses the current TMDP and PLDP mechanisms~\cite{nissim2011distributed,Murakami019}.

\subsection{Algorithmic Description}

Different from the general function $f(\cdot) = \{f_i(\cdot)\}_{i\in [n]}$ considered in MPC, where the functions $f_i(\cdot),\, i\in [n]$ can be arbitrarily different, in the multi-party Gaussian mechanism,
we consider a less general form of $f$, where $f_i(\cdot)$ is either equal to $F(\cdot)$ or $\perp$ for $i\in [n]$.
Here, $F(\cdot)$ represents an arbitrary function, and $\perp$ is a symbol with no information. Thus, $f$ computes $F$ and reveals the result to a subset of parties while outputting $\perp$ to the others. This idea is similar to that adopted in MPC for defining protocols where the output is revealed only to a subset of parties. Note that there is no loss of generality since a general function can be composed of a finite number of such less general functions, which satisfies a TPMDP requirement (see the composition theorem in
Appendix~\ref{app:comp}).

To simplify notation, we define the set of active users as ${U^+} = \{i \in [n] | f_i ({\boldsymbol{x}}) = F({\boldsymbol{x}})\}$, where the non-active set is the complement of ${U^+}$ and is denoted by ${U^-}$. Additionally, we define the active adversary structure ${\mathcal{A}_{t}^+}$ as the set of adversarial sets containing at least one active user, that is, $\mathcal{A}_{t}^+ = \{A\in \mathcal{A}_t | A\cap U^+ \neq \emptyset\}$.

The general algorithm of the multi-party Gaussian mechanism is presented in Alg.~\ref{alg:DP_Gaussian}. This algorithm perturbs the query results by adding the sum of all parties' Gaussian noises in an MPC protocol for $f$.

\begin{algorithm}
\setstretch{0.8}
    \caption{Multi-party Gaussian mechanism}
    \label{alg:DP_Gaussian}
    \begin{algorithmic}[1]
        \STATE {\bfseries Input:} {$F(\cdot)$, $(t,n,\mathcal{E})$, ${\boldsymbol{x}}$, ${U^+}$}
        \STATE Each party $i$ computes a noise variance $\sigma_i^2$ 
        \STATE Each party $i$ generates $X_i \sim \mathcal{N}(0,\sigma_i^2 I)$
        \STATE All parties execute an MPC protocol $\pi$ with inputs ${\boldsymbol{x}}$, ${\boldsymbol{X}}$ and an output $\Pi_i({\boldsymbol{x}},{\boldsymbol{X}}) = F({\boldsymbol{x}})+\Sigma_{j=1}^n X_j$ to $i \in {U^+}$ and $\perp$ to $i\in U^-$, {where $\Pi_i$ denotes the function w.r.t the $i$-th output of $\Pi$}
    \end{algorithmic}
\end{algorithm}
The multi-party Gaussian mechanism will not result in significant communication overhead to the original MPC protocol since only $n$ addition gates need to be included.
For instance, if the traditional Shamir secret-sharing scheme~\cite{goldreich2007foundations} is employed, this can be accomplished in $\mathcal{O} (1)$ extra rounds with an $\mathcal{O}(n)$ additional message complexity to each party.

\hidemass{
In a finite-precision implementation, the Gaussian distribution in Line $2$ of Alg.~\ref{alg:DP_Gaussian} can be substituted with a discrete Gaussian distribution that has been sufficiently discretized. The formal definition for the discrete Gaussian distribution is provided in the full version. By setting the discretization step as $2^{-\Theta(\left|{\boldsymbol{x}}\right|)}$, the cumulative distribution function (CDF) of the discrete Gaussian distribution converges pointwise exponentially fast to that of the continuous Gaussian distribution as $\left|{\boldsymbol{x}}\right|$ increases
\cite{Canonne0S20}.
This fact can be used to establish that the privacy property provided by the sum of independent discrete Gaussian noises with sufficient discretization is essentially the same as that provided by a single discrete Gaussian noise whose variance parameter equals the sum of the variances of the previous ones. We formalize this property in the full version. Consequently, the analysis of privacy and utility, in this case, follows a similar approach to that for the Gaussian distribution.
}

\subsection{Privacy Analysis and Optimization Objective}\label{subsec:privacy-opt-objective}



This section discusses the conditions under which the multi-party Gaussian mechanism satisfies TPMDP. To maximize the utility of the mechanism while ensuring TPMDP, we formulate a variance-minimization problem.

We introduce the concept of partial $\ell_2$-sensitivity, which measures the maximum difference in outcomes of a query function on neighboring inputs (cf. Definition~\ref{def:p_l2_sensitivity}). This sensitivity value is essential in determining the appropriate noise variance to ensure TPMDP.


\begin{definition}[Partial $\ell_2$-sensitivity]\label{def:p_l2_sensitivity}
    Given a function $F: (\{0,1\}^*)^n\rightarrow \mathbb{R}^d$, the $i$-th partial $\ell_2$-sensitivity of $F$ is 
    $$\Delta_{2,i} F = \mathop{\max}\limits_{\mathop{x,y\in{(\{0,1\}^*)^n}}, x \overset{i}{\simeq} y} \left\|F(x)-F(y)\right\|_2.$$
\end{definition}

We demonstrate that the privacy guarantee of the multi-party Gaussian mechanism can be assessed by comparing partial sums of $\{\sigma_i^2\}_{i\in [n]}$ and $\{\sigma_{\Gamma_i}^2\}_{i\in [n]}$, where
$\sigma_{\Gamma_i}^2$ denotes the sufficient noise variance for Gaussian mechanisms with parameters $\Gamma_i \triangleq (\epsilon_i, \delta_i, \Delta_{2,i} F)$.
The theorem for the multi-party Gaussian mechanism is established
as follows.
We defer the proof of Theorem~\ref{thm:DP_multi_Gaussian_stat} to
Appendix~\ref{app:prob}.

\begin{theorem}
    \label{thm:DP_multi_Gaussian_stat}

    A multi-party Gaussian mechanism $\Pi$ satisfies statistical $(t,n,\mathcal{E})$-TPMDP if the MPC protocol $\pi$ in Alg.~\ref{alg:DP_Gaussian} is statistically $\tau$-private where $\tau \geq t$ and $\boldsymbol{\sigma}$ satisfies
    \begin{equation}\label{eq:DP_multi_Gaussian_stat}
        \Sigma_{i\in \bar{A_j}} \sigma_i^2 \geq \sigma_{\Gamma_j}^2,\  \forall j\in [n], \forall A_j \in \mathcal{A}_{t}^+ \cap \{A | j\in \bar{A}, |A| = t\},
    \end{equation}
    %
    where $\bar{A}$ denotes the complement of the set $A$.
    $\Pi$ satisfies computational $(t,n,\mathcal{E})$-TPMDP if
    $\pi$ is computationally $\tau$-private where $\tau \geq t$ and Eq.~\eqref{eq:DP_multi_Gaussian_stat} holds.
\end{theorem}

The theory indicates that a specific multi-party Gaussian mechanism can satisfy a given TPMDP requirement with parameters $n$, $t$, and $\mathcal{E}$. To optimize the utility of the multi-party Gaussian mechanism, we aim to minimize the additive noise variance while satisfying the TPMDP constraints given by Eq.~\eqref{eq:DP_multi_Gaussian_stat}. This involves finding each party's variance $\sigma_i^2$ by solving the following optimization problem:
\begin{equation}\label{eq:Gauss_opt}
    \begin{aligned}
        &\min_{\boldsymbol{\sigma}=\{\sigma_i\}_{i=1}^n} \Sigma_{i=1}^n \sigma_i^2 \\
        \Sigma_{i\in \bar{A_j}} \sigma_i^2 \geq \sigma_{\Gamma_j}^2,\  &\forall j\in [n], \forall A_j \in \mathcal{A}_{t}^+ \cap \{A | j\in \bar{A}, |A| = t\}.
    \end{aligned}
\end{equation}

The optimization problem described in  Eq.~\eqref{eq:Gauss_opt} has noise variances $\sigma_i^2, i\in [n]$ as decision variables. This indicates that the problem can be classified as a linear programming (LP) problem. For the sake of convenience, we define the objective function in Eq. \eqref{eq:Gauss_opt} as $v_{\boldsymbol{\sigma}}$, such that $v_{\boldsymbol{\sigma}} := \sum_{i=1}^n \sigma_i^2$.

\subsection{Utility-Maximizing Algorithm}\label{subsec:utility-max-alg}

In this section, we present an efficient method to obtain a utility-optimal multi-party Gaussian mechanism by solving the optimal noise variances in Eq.~\eqref{eq:Gauss_opt}.
Eq.~\eqref{eq:Gauss_opt} is an LP problem, which can be solved using a generic solver to obtain the optimal mechanism. The current state-of-the-art complexity of generic solvers for LP problems is $\widetilde{\mathcal{O}}((nnz+rank^2)\sqrt{rank}\log{(1/\epsilon)})$~\cite{lee2019solving}, where $nnz$ and $rank$ refer to the number of non-zero entries and rank of the constraint matrix, respectively, and $\epsilon$ is the approximation parameter. The complexity of solving Eq.~\eqref{eq:Gauss_opt} is polynomial in $n$, with the order depending on $t$.
\comment{This complexity limits the scalability of the multi-party Gaussian mechanism for large values of $t$.}
To overcome this limitation, we propose a method that solves Eq.~\eqref{eq:Gauss_opt} exactly, i.e., obtaining an exact solution in a finite number of iterations.
Specifically, our method requires only $\mathcal{O}(n)$ computations.
We provide a detailed comparison between the complexity of the generic LP solver and our method at the end of this subsection.
%
Besides,
the utility of our multi-party Gaussian mechanism, as measured by the noise variance, is superior to both TMDP and PLDP mechanisms~\cite{nissim2011distributed,Murakami019}.

We present our results for two cases. First, we consider a special case where ${U^+}= [n]$ (i.e., all parties are active), which implies ${\mathcal{A}_{t}^+} = \mathcal{A}_{t}$. We then extend our results to the general case where ${U^+}\subset [n]$. We exclude two trivial cases from our analysis, where ${U^+} = \emptyset$ and $t = 0$, as in both cases, ${\mathcal{A}_{t}^+} = \emptyset$, which means that no noise is required. Therefore, we focus on $1\leq \left|U^+ \right|\leq n$ and $1\leq t\leq n-1$ in the subsequent analysis. The proofs for all results presented in this section can be found in
Appendix~\ref{app:res-pf}.

\hidemass{
At a high level,
when $U^+ = [n]$,
our algorithm proceeds as follows:
1) Each party computes sufficient variances for all parties' DP requirements;
2) The parties are divided into two groups, stringent and non-stringent, based on their required noise level;
3) A uniform noise variance is assigned to the parties to achieve the non-stringent parties' privacy requirements;
4) Additional noise variances are assigned to the stringent parties if the uniform noise variance is not sufficient.
For
a general $U^+\subset [n]$, whether the output is revealed to a party is also crucial. For instance, if the output is only revealed to a single party, then that party
does not need to add noise.
}

\vspace{5pt}
\noindent \emph{\textbf{1) Case 1: ${U^+} = [n]$}}
\vspace{3pt}
%





Regarding Case 1, based on a dedicated analysis of the structure of the LP problem in Eq.~\eqref{eq:Gauss_opt}, we discover that the optimal solution can be represented concisely. Specifically, the solution can be found as follows: 1) Each party computes sufficient variances for all parties' DP requirements; 2) A uniform noise variance is assigned to the parties to satisfy the majority of the privacy requirements; 3) Additional noise variance is assigned to a party if the uniform noise variance is insufficient to meet his/her privacy requirement.

\comment{For ease of elaboration, we denote $\sigma_{\Gamma_{(i)}}$ as the $i$-th largest element in $\{\sigma_{\Gamma_j}\}_{j\in [n]}$.}
The exact method for computing the optimal $\sigma_i^2,\ i\in [n]$ for Eq.~\eqref{eq:Gauss_opt} in Case $1$ is synopsized in Alg. \ref{alg:DP_OPS_trivial}.
Specifically, after computing $\xi$ in $\mathcal{O}(1)$ steps, the computation of $\sigma_i^2$ can be broken down into two parts. The first part involves finding the $\xi$-th largest element in $\{\sigma_{\Gamma_j}\}_{j\in [n]}$, which has a complexity of $\mathcal{O}(n)$. The second part is a closed-form expression with a complexity of $\mathcal{O}(1)$. Therefore, the overall complexity is $\mathcal{O}(n)$.

\begin{algorithm}
\setstretch{0.8}
    \caption{Optimal parameter selection for multi-party Gaussian mechanism when ${U^+} = [n]$}
    \label{alg:DP_OPS_trivial}
    \begin{algorithmic}[1]
        \STATE {\bfseries Input: }{$F$, $(t,n,\mathcal{E})$, $i$}
        \STATE {\bfseries Output: }{Optimal $\sigma_i^2$ for party $i$ in Eq.~\eqref{eq:Gauss_opt}}
            \STATE Compute $\sigma_{\Gamma_j}$ for $j\in [n]$ and $\xi = \min{(\lfloor \frac{2n-t}{n-t} \rfloor, t+1)}$\\
            \STATE Find the $\xi$-th largest element in $\{\sigma_{\Gamma_j}\}_{j\in [n]}$, $\sigma_{\Gamma_{(\xi)}}$ \\
            \STATE Set $\sigma_i^2 = {\frac{1}{n-t} \sigma_{\Gamma_{(\xi)}}^2}$ if $\sigma_{\Gamma_i}\leq \sigma_{\Gamma_{(\xi)}}$ and $\sigma_i^2 = {\sigma_{\Gamma_i}^2 - \frac{n-t-1}{n-t} \sigma_{\Gamma_{(\xi)}}^2}$ if $\sigma_{\Gamma_i} > \sigma_{\Gamma_{(\xi)}}$ \\
    \end{algorithmic}
\end{algorithm}

\hide{
\begin{algorithm}
 \SetAlgoVlined
{
    \caption{Optimal parameter selection for multi-party Gaussian mechanism when ${U^+} = [n]$}
    \label{alg:DP_OPS_trivial}
    \SetKwInOut{Input}{Input}\SetKwInOut{Output}{Output}
    \Input{$F$, $(t,n,\mathcal{E})$, $i$}
    \Output{Optimal $\sigma_i^2$ for party $i$ w.r.t Eq.~\eqref{eq:Gauss_opt}}
    \If{$t=n-1$}{\
    Compute $\sigma_{\Gamma_i}$\\
    Set $\sigma_i^2 = \sigma_{\Gamma_i}^2$\\
    }
    \Else
    {
        Compute $\sigma_{\Gamma_j}$ for $j\in \{1,\hdots,n\}$\\
        Set $\xi = \min{(\lfloor \frac{2n-t}{n-t} \rfloor, t+1)}$\\
        Find the $\xi$-th largest element in $\{\sigma_{\Gamma_j}\}_{j=1}^n$, denoted by $\sigma_{\Gamma_{(\xi)}}$ \\
        Set $\sigma_i^2 = {\frac{1}{n-t} \sigma_{\Gamma_{(\xi)}}^2}$ if $\sigma_{\Gamma_i}\leq \sigma_{\Gamma_{(\xi)}}$ and $\sigma_i^2 = {\sigma_{\Gamma_i}^2 - \frac{n-t-1}{n-t} \sigma_{\Gamma_{(\xi)}}^2}$ if $\sigma_{\Gamma_i} > \sigma_{\Gamma_{(\xi)}}$ \\
    }
    }
\end{algorithm}
}

\begin{theorem}\label{thm:case_trivial}
    For Case 1, Alg. \ref{alg:DP_OPS_trivial} outputs the optimal entries $\sigma_i^2,\ i\in [n]$ for Eq.~\eqref{eq:Gauss_opt}, where the computation takes
    $\mathcal{O}(n)$ steps. 
    \comment{The optimal
    overall noise variance is
    $v_{\boldsymbol{\sigma}} = \Sigma_{i=1}^{\xi-1} \sigma_{\Gamma_{(i)}}^2 + (\frac{2n - t}{n - t} - \xi) \sigma_{\Gamma_{(\xi)}}^2$, where $\xi = \min{(\lfloor \frac{2n-t}{n-t} \rfloor, t+1)}$}
\end{theorem}


Theorem~\ref{thm:case_trivial} demonstrates that in the case where $U^+ = [n]$, each party can determine the optimal variance with a complexity of $\mathcal{O}(n)$.
The theorem further shows that when the threshold $t$ is relatively small, i.e., $t \leq p n$ for some constant $p<1$, the value of $\xi$ is bounded by $\mathcal{O}(1)$. In this scenario, the overall noise variance $v_{\boldsymbol{\sigma}}$ is $\mathcal{O}(1)$ and is independent of $n$. However, if the threshold $t$ is close to $n$ (e.g., $n - t = \mathcal{O}(1)$), the value of $\xi$ scales as $\mathcal{O}(n)$. Consequently, the overall noise variance is essentially the sum of required noise variances of a constant percentage of learners with more stringent privacy requirements.


\vspace{5pt}
\noindent \emph{\textbf{2) Case 2: ${U^+} \subset [n]$}}
\vspace{3pt}




In this subsection, we extend our analysis to the general case where $U^+\subset [n]$. We assume that $U^+$ has size $\eta+1$, where $0\leq \eta \leq n-1$, and for convenience, we reparameterize $U^+$ as $\{1^+,2^+,\hdots,(\eta+1)^+\}$ and ${U^-}$ as $\{1^-,\hdots,(n-\eta-1)^-\}$. Let ${\mathcal{E}^+}$ and ${\mathcal{E}^-}$ be the sets of pairs $(\epsilon_i,\delta_i)$ for $i\in U^+$ and $i\in U^-$, respectively, and let $\boldsymbol{\sigma}^+$ and $\boldsymbol{\sigma}^-$ be the sets of $\sigma_i$ for $i\in U^+$ and $i\in U^-$, respectively.

As before, we define $\sigma_{\Gamma_{(i^+)}}$ and $\sigma_{\Gamma_{(i^-)}}$ as the $i$-th largest element in $\{\sigma_{\Gamma_j}\}_{j\in U^+}$ and $\{\sigma_{\Gamma_j}\}_{j\in U^-}$, respectively.
The results are classified into four mutually exclusive subcases:
\begin{enumerate}\label{subcase}
    \item \textbf{Subcase 1}\label{subcase_1}: $\left| {U^+} \right| \geq n-t+1$;
    \item \textbf{Subcase 2}\label{subcase_2}: $\left| {U^+} \right| = 1$;
    \item \textbf{Subcase 3}\label{subcase_3}: $2 \leq \left| {U^+} \right| \leq n-t$ and $n-t \left| {U^+} \right| \leq 0$;
    \item \textbf{Subcase 4}\label{subcase_4_ext}: $2 \leq \left| {U^+} \right| \leq n-t$ and $n-t \left| {U^+} \right| > 0$.
\end{enumerate}
These subcases cover all possibilities for
$U^+$ and $t$.

\begin{lemma}\label{lemma:DP_active_equiv}
    For Subcases \ref{subcase_1} and \ref{subcase_3}, $\boldsymbol{\sigma}$ is optimal for Eq.~\eqref{eq:Gauss_opt} if it is optimal for Eq.~\eqref{eq:Gauss_opt} when replacing ${\mathcal{A}_{t}^+}$ with ${\mathcal{A}_{t}}$.
\end{lemma}


According to Lemma \ref{lemma:DP_active_equiv}, determining the optimal value of $\boldsymbol{\sigma}$ for Subcases \ref{subcase_1} and \ref{subcase_3} is equivalent to finding the optimal value of $\boldsymbol{\sigma}$ after substituting ${\mathcal{A}_{t}^+}$ with ${\mathcal{A}_{t}}$. This task can be accomplished using Alg. \ref{alg:DP_OPS_trivial}.


\begin{lemma}\label{lemma:{U^+}_1_trivial}
    For Subcase \ref{subcase_2}, $\boldsymbol{\sigma}$ is optimal for Eq.~\eqref{eq:Gauss_opt} if:
    \begin{enumerate}
        \item when $t \geq 2$, $\sigma_{{1^+}}^2 = 0$ and ${\boldsymbol{\sigma}^-}$ is optimal for Eq.~\eqref{eq:Gauss_opt} with input $(t-1,n-1,{\mathcal{E}^-})$ and active set ${U^-}$;
        \item when $t = 1$, $\sigma_{{1^+}}^2 = 0$, $\sigma_i^2 = \frac{1}{n-1} \sigma_{\Gamma_{(1^-)}}^2$ for $i\in U^-$.
    \end{enumerate}
\end{lemma}

\begin{lemma}\label{lemma:nontrivial_noise_active_ext}
    For Subcase \ref{subcase_4_ext},
    let $\alpha = \max\{\sigma_{\Gamma_{(1^-)}}, \sigma_{\Gamma_{(2^+)}}\}$ and $\beta = \max \{\sigma_{\Gamma_{(1^+)}}, \sigma_{\Gamma_{(2^-)}}\}$.
    Then $\boldsymbol{\sigma}$ is optimal for Eq.~\eqref{eq:Gauss_opt}
    if:
    \begin{enumerate}
        \item when $t=1$ or $\alpha \leq \beta$, $\sigma_i^2 = \frac{1}{n-\eta-t} \alpha^2$ for $i\in {U^-}$,
        and $\sigma_{{i}}^2 = \max\{0,\sigma_{\Gamma_{{i}}}^2 - \alpha^2\}$ for $i\in {U^+}$;
        %
        %
        \item when $t\geq 2$ and $\alpha > \beta$,
        $\sigma_i = 0$ if $i\in U^+$,
        $\sigma_{i}^2 = \alpha^2 - \frac{n-\eta-t-1}{n-\eta-t} \beta^2$
        if $i\in U^-$ and $\sigma_{\Gamma_i} = \sigma_{\Gamma_{(1^-)}}$,
        and
        $\sigma_i^2 = \frac{1}{n-\eta-t} \beta^2$ otherwise.
    \end{enumerate}
\end{lemma}

Alg. \ref{alg:DP_OPS_2} outlines the complete procedure for all four subcases. Within this algorithm, each learner identifies the subcase under which the parameter setting falls, then calculates the noise variance using the corresponding lemma described above. Like Alg.~\ref{alg:DP_OPS_trivial}, the computational complexity of Alg. \ref{alg:DP_OPS_2} is primarily determined by the process of finding $\sigma_{\Gamma_i}$ for $i\in [n]$ and the $\xi$-th largest element for some $\xi \in [n]$. Consequently, the complexity of Alg. \ref{alg:DP_OPS_2} is $\mathcal{O} (n)$.

\begin{algorithm}[tp]
\setstretch{0.8}
\caption{Optimal parameter selection for multi-party Gaussian mechanism}
\label{alg:DP_OPS_2}
\begin{algorithmic}[1]
    \STATE {\bfseries Input: }{$F$, $(t,n,\mathcal{E})$, ${U^+}$, $i$}
    \STATE {\bfseries Output: }{Optimal $\sigma_i^2$ for party $i$ in Eq.~\eqref{eq:Gauss_opt}}
    \STATE Compute $\sigma_{\Gamma_j}$ for $j\in [n]$ and $\xi = \min{(\lfloor \frac{2n-t}{n-t} \rfloor, t+1)}$\\
    \STATE Find the largest and $2$-nd largest elements in $\{\sigma_{\Gamma_j}\}_{j\in {U^-}}$, $\sigma_{\Gamma_{(1^-)}}$ and $\sigma_{\Gamma_{(2^-)}}$
    \STATE Find the largest and $2$-nd largest element in $\{\sigma_{\Gamma_j}\}_{j\in {U^+}}$, $\sigma_{\Gamma_{(1^+)}}$ and $\sigma_{\Gamma_{(2^+)}}$
    \STATE Set {\small{$\alpha = \max\{\sigma_{\Gamma_{(1^-)}}, \sigma_{\Gamma_{(2^+)}}\}$}} and {\small{$\beta = \max \{\sigma_{\Gamma_{(1^+)}}, \sigma_{\Gamma_{(2^-)}}\}$}}
    \SWITCH{Subcase}
        \CASE{Subcase \ref{subcase_1}, Subcase \ref{subcase_3}}
            \STATE Compute $\sigma_{i}^2$ using Alg. \ref{alg:DP_OPS_trivial} with input {\small{$(F, (t, n, \mathcal{E}), i)$}}
            \STATE Break
        \ENDCASE
        \CASE{Subcase \ref{subcase_2}}
            \IF{$i\in {U^+}$}
                \STATE Set $\sigma_{i}^2 = 0$
            \ELSIF{$t\geq 2$}
                \STATE Compute $\sigma_{i}^2$ via Alg. \ref{alg:DP_OPS_trivial} with \hidemass{input} {\small{$(F, (t-1, n-1, {\mathcal{E}^-}), i)$}}
            \ELSE
                \STATE Set $\sigma_i^2 = {\frac{1}{n-1} \sigma_{\Gamma_{(1^-)}}^2}$
            \ENDIF
            \STATE Break
        \ENDCASE
        \CASE{Subcase \ref{subcase_4_ext}}
            \IF{$t = 1$ or $\alpha \leq \beta$}
                \IF{$i\in {U^+}$}
                    \STATE Set $\sigma_i^2 = \max\{0,\sigma_{\Gamma_{{i}}}^2 - \alpha^2\}$
                \ELSE
                    \STATE Set $\sigma_i^2 = {\frac{1}{n-\left|{U^+} \right|-t +1} \alpha^2}$\\
                \ENDIF
            \ELSE
                \IF{$i\in {U^-}$ and $\sigma_{\Gamma_i} = \sigma_{\Gamma_{(1^-)}}$}
                    \STATE Set $\sigma_i^2 = {\alpha^2 - \frac{n-\left|{U^+} \right|-t}{n-\left|{U^+} \right|-t+1} \beta^2}$\\
                
                \ELSIF{$i\in {U^-}$}{
                    \STATE Set {$\sigma_i^2 = {\frac{1}{n-\left|{U^+} \right|-t+1} \beta^2}$}\\
                }
                \ELSE{
                    \STATE Set $\sigma_i^2 = 0$\\
                }
                \ENDIF
            \ENDIF
        \ENDCASE
    \ENDSWITCH
\end{algorithmic}
\end{algorithm}



\begin{theorem}\label{thm:case_general_rst}
    Alg. \ref{alg:DP_OPS_2} outputs the optimal
    $\sigma_i^2,\ i\in [n]$
    for Eq.~\eqref{eq:Gauss_opt}. The computation takes
    $\mathcal{O}(n)$ steps.
%
    The optimal
    overall noise variance is
    \begin{equation}
    v_{\boldsymbol{\sigma}} =
    \left\{
        \begin{aligned}
            & \Sigma_{i=1}^{\xi-1} \sigma_{\Gamma_{(i)}}^2 + \Big(\frac{2n - t}{n - t} - \xi\Big) \sigma_{\Gamma_{(\xi)}}^2,\qquad Subcases~\ref{subcase_1}\&\ref{subcase_3} \\
            & \Sigma_{i=1}^{\xi-1} \sigma_{\Gamma_{(i^-)}}^2 + \Big(\frac{2n - t - 1}{n - t} - \xi\Big) \sigma_{\Gamma_{(\xi^-)}}^2,\ \ Subcase~\ref{subcase_2} \\
            & \sigma_{\Gamma_{(1)}}^2 + \frac{t - 1}{n - \left| U^+ \right| - t +1} \sigma_{\Gamma_{(2)}}^2,\qquad\qquad Subcase~\ref{subcase_4_ext}, \\
        \end{aligned}
    \right.
    \end{equation}
    where $\xi= \min{(\lfloor \frac{2n-t}{n-t} \rfloor, t+1)}$ in Subcases~\ref{subcase_1} and \ref{subcase_3}, and $\xi= \min{(\lfloor \frac{2n-t - 1}{n-t} \rfloor, t)}$ in Subcase~\ref{subcase_2}.
\end{theorem}


Theorem~\ref{thm:case_general_rst} demonstrates that each learner can achieve the optimal variance with a complexity of $\mathcal{O}(n)$ for the general case where $U^+\subset [n]$.
Regarding the overall noise variance $v_{\boldsymbol{\sigma}}$, if the threshold $t\leq pn$ for a constant $p\leq1$, similar to the case $U^+= [n]$, it holds that $v_{\boldsymbol{\sigma}}=\mathcal{O}(1)$ for Subcases \ref{subcase_1}-\ref{subcase_4_ext} when considering the dependence of $n$. However, when $t$ is close to $n$ (e.g., $n-t=\mathcal{O}(1)$), the parameter settings in Subcase \ref{subcase_4_ext} do not exist. In Subcases \ref{subcase_1}-\ref{subcase_3}, $\xi=O(n)$, and the overall noise variance essentially equals the sum of the required noise variances of a constant percentage of learners with more stringent privacy budgets.

\textbf{Comparison with existing works}: Table~\ref{tab:complexity} summarizes the space and time complexity comparisons between Alg.~\ref{alg:DP_OPS_2} and a generic LP solver for Eq.~\eqref{eq:Gauss_opt}. The results demonstrate that, when $t = p \cdot n$ for $p<1$, Alg.~\ref{alg:DP_OPS_2} reduces the space and time complexities of the generic solver from exponential to $\mathcal{O}(n)$.

Regarding utility, the previous TMDP mechanism~\cite{nissim2011distributed} did not consider personalized privacy requirements for each party.
In a multi-party Gaussian mechanism satisfying TMDP, we replace each party $i$'s required noise variance $\sigma_{\Gamma_i}^2$ with the largest of their required noise variances $\sigma_{\Gamma_{(1)}}^2$.
Compared to our TPMDP multi-party Gaussian mechanism, the overall noise variance $v_{\boldsymbol{\sigma}}$ can be prohibitively large if the threshold value is close to $n$ and all parties' privacy requirements are heterogeneous. For example, if $\sigma_{\Gamma_{(2)}} \ll \sigma_{\Gamma_{(1)}}$ and $n - t = \mathcal{O}(1)$, it holds that $v_{\boldsymbol{\sigma}}^{\text{TPMDP}} / v_{\boldsymbol{\sigma}}^{\text{TMDP}} \leq \mathcal{O}(1) \cdot \max\{\frac{1}{n}, \frac{\sigma_{\Gamma_{(2)}}^2}{\sigma_{\Gamma_{(1)}}^2}\} \ll 1$ for sufficiently large $n$ in Subcases~\ref{subcase_1} and \ref{subcase_3}. Here, $v_{\boldsymbol{\sigma}}^{\text{TPMDP}}$ and $v_{\boldsymbol{\sigma}}^{\text{TMDP}}$ denote the optimal variance sums of TPMDP and TMDP multi-party Gaussian mechanisms, respectively.
Compared to the PLDP mechanism~\cite{kairouz2015secure}, our mechanism has a prominent advantage when $t \leq pn$. In this case, the overall noise variances of PLDP and our mechanisms scale as $\mathcal{O}(n)$ and $\mathcal{O}(1)$, respectively, provided that $\sigma_{\Gamma_i} \in [C_1, C_2]$ for $i \in [n]$ and $0 < C_1 < C_2$.


\begin{table}[htbp]
\setstretch{0.8}
\caption{
Complexity for generic LP solver and Alg.~\ref{alg:DP_OPS_2}.
$H(p) = p \log{p} + (1-p) \log{(1-p)}$ is the entropy for $p < 1$.
The space complexity is evaluated using the size of codes and inputs the algorithm requires.
The time complexity for the generic LP solver is
from the result in \cite{lee2019solving} (i.e., $\widetilde{\mathcal{O}}((nnz+rank^2)\sqrt{rank}\log{(1/\epsilon)})$, where
$nnz$ and $rank$ denote the number of non-zero entries and rank of the constraint matrix,
and $\epsilon$ measures the suboptimality of the LP solver),
after a plain application of Stirling's formula.
}\label{tab:complexity}
{\small
\begin{tabular}{ >{\centering\arraybackslash} m{2.25cm}  >{\centering\arraybackslash} m{2.47cm}  >{\centering\arraybackslash} m{2.7cm} }
    \toprule
     & Space complexity & Time complexity \\
    \midrule
    Generic solver ($t = C$) & $\mathcal{O}(n^{C+2})$ & $\widetilde{\mathcal{O}}(n^{C+2.5}) \log{(1/\epsilon)}$ \\
    Generic solver ($t = p\cdot n$) & $\mathcal{O}(n^{1.5} 2^{H(p)n})$ & $\widetilde{\mathcal{O}}(n^{2} 2^{nH(p)}) \log{(1/\epsilon)}$ \\
    Alg. \ref{alg:DP_OPS_2} & $\mathcal{O}(n)$ & $\mathcal{O}(n)$ \\
    \bottomrule
\end{tabular}
}
\end{table}

\hide{
\begin{algorithm}[tp]
\setstretch{0.7}
{
  \caption{Optimal parameter selection for multi-party Gaussian mechanism}
    \label{alg:DP_OPS_2}
     \SetKwInOut{Input}{Input}\SetKwInOut{Output}{Output}
    \Input{$F$, $(t,n,\mathcal{E})$, ${U^+}$, $i$}
    \Output{Optimal $\sigma_i^2$ for party $i$ w.r.t Eq.~\eqref{eq:Gauss_opt}}
    \If{$\left| {U^+} \right| = 1$}{
        \If{$i\in {U^+}$}{
            Set $\sigma_{i}^2 = 0$\\
        }
        \ElseIf{$t\geq 2$}{
            Compute $\sigma_{i}^2$ using Alg. \ref{alg:DP_OPS_trivial} with parameter $(F, (t-1, n-1, {\mathcal{E}^-}), i)$\\
        }
        \Else
        {
            Find the largest element in $\{\sigma_{\Gamma_j}\}_{j\in {U^-}}$, denoted by $\sigma_{\Gamma_{(1)}}$\\
            Set $\sigma_i^2 = {\frac{1}{n-1} \sigma_{\Gamma_{(1)}}^2}$\\
        }
   }
   \Return \\

    \If{$\left | {U^+} \right | \geq n-t+1$ or $n-t\left |{U^+} \right | \leq 0$}{
        Compute $\sigma_{i}^2$ using Alg. \ref{alg:DP_OPS_trivial} with parameter $(F, (t, n, \mathcal{E}), i)$\\
    }
    \Return \\
    
    Compute $\sigma_{\Gamma_j}$ for $j\in [n]$\\
    Find the largest and the $2$-largest element, $\sigma_{\Gamma_{({1^+})}}$ and $\sigma_{\Gamma_{(2^*)}}$ in $\{\sigma_{\Gamma_j}\}_{j\in {U^+}}$\\
    Find the largest and the $2$-largest element, $\sigma_{\Gamma_{(1)}}$ and $\sigma_{\Gamma_{(2)}}$ in $\{\sigma_{\Gamma_j}\}_{j\in {U^-}}$\\

    Set $\overline{\sigma_{\Gamma_{(1)}}} = \max\{\sigma_{\Gamma_{(1)}}, \sigma_{\Gamma_{(2^*)}}\}$; Set $\overline{\sigma_{\Gamma_{(2)}}} = \max\{\sigma_{\Gamma_{(2)}}, \sigma_{\Gamma_{(2^*)}}\}$\\
    
    \If{$t = 1$}{
        \If{$i\in {U^+}$ and $\sigma_{\Gamma_i} = \sigma_{\Gamma_{({1^+})}}$}{
            Set $\sigma_i^2 = {\max \{0, \sigma_{\Gamma_{({1^+})}}^2 - \overline{\sigma_{\Gamma_{(1)}}}^2\}}$\\
        }
        \ElseIf{$i\in {U^+}$}{        
            Set $\sigma_i^2 = 0$\\
        }
        \Else{ Set $\sigma_i^2 = {\frac{1}{n-\left|{U^+} \right|} \overline{\sigma_{\Gamma_{(1)}}}^2}$\\
        }
    }
    \Return \\
        
         \If{$\overline{\sigma_{\Gamma_{(1)}}} \leq \max \{\sigma_{\Gamma_{({1^+})}}, \overline{\sigma_{\Gamma_{(2)}}}\}$}{
            \If{$i\in {U^+}$ and $\sigma_{\Gamma_i} = \sigma_{\Gamma_{({1^+})}}$}{
                Set $\sigma_i^2 = {\max\{0, \sigma_{\Gamma_{({1^+})}}^2 - \overline{\sigma_{\Gamma_{(1)}}}^2\}}$\\
            }
            \ElseIf{$i\in {U^+}$}{
                Set $\sigma_i^2 = 0$\\
            }
            \Else{
                Set $\sigma_i^2 = {\frac{1}{n-\left|{U^+} \right|-t +1} \overline{\sigma_{\Gamma_{(1)}}}^2}$\\
            }
        }
        \Else{
            \If{$i\in {U^-}$ and $\sigma_{\Gamma_i} = \sigma_{\Gamma_{(1)}}$}{
                Set $\sigma_i^2 = {\overline{\sigma_{\Gamma_{(1)}}}^2 - \frac{n-\left|{U^+} \right|-t}{n-\left|{U^+} \right|-t+1} \max\{ \sigma_{\Gamma_{({1^+})}}^2, \overline{\sigma_{\Gamma_{(2)}}}^2\}}$\\
            }
            \ElseIf{$i\in {U^-}$}{
                Set {$\sigma_i^2 = {\frac{1}{n-\left|{U^+} \right|-t+1} \max\{ \sigma_{\Gamma_{({1^+})}}^2, \overline{\sigma_{\Gamma_{(2)}}}^2\}}$}\\
            }
            \Else{
                Set $\sigma_i^2 = 0$\\
            }
        }
    }
\end{algorithm}
}


\section{Evaluations by Experiments}
\label{sec:perform}

This section assesses the utility and scalability of our multi-party Gaussian mechanism. Our findings demonstrate that our mechanism surpasses the heuristic non-threshold approach and the PLDP method in terms of utility and is even comparable to a centralized approach. Furthermore, our mechanism achieves better utility than the TMDP mechanism, particularly when the threshold value $t$ approaches $n$ or the proportion of conservative parties is relatively low. Additionally, our algorithm (Alg.~\ref{alg:DP_OPS_2}) significantly outperforms the leading generic LP solver concerning running time and storage requirements.

\begin{figure*}[h!]
\centering
\includegraphics[width=\linewidth]{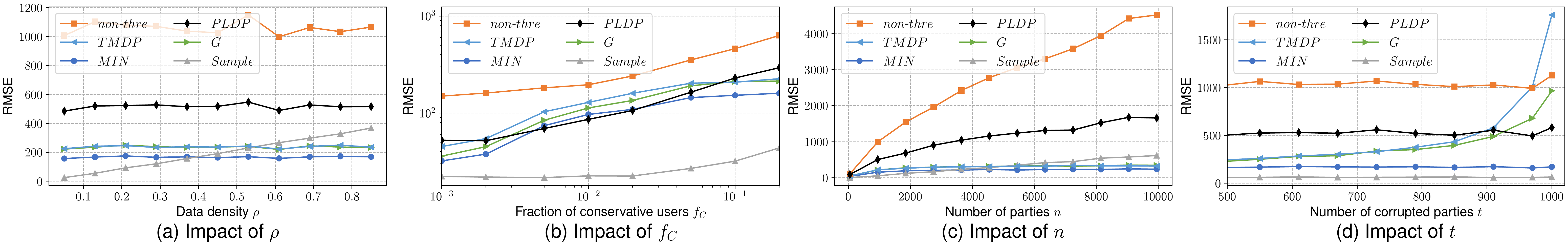}
\caption{Utility loss for count.
$G$ is our proposed mechanism; \emph{MIN} and \emph{non-thre} are centralized and non-threshold baselines, respectively; \emph{Sample}, \emph{PLDP}, and \emph{TMDP} stand for the Sample, randomized response, and TMDP mechanisms.}\label{fig:cnt}
\vspace{-0.1in}
\end{figure*}

\subsection{Utility Loss}

\textbf{Experimental setup}: We evaluate the performance of our proposed mechanism by measuring the utility loss in two common query functions: count and linear regression. Count is a basic operation in complex data analytic tasks, while linear regression is a fundamental machine learning algorithm. We measure the utility loss of the mechanism using rooted mean squared error (RMSE). Additionally, we compare our proposed multi-party Gaussian mechanism
(referred to as $G$ in this section)
with several baseline methods:
\begin{itemize}
    \item \textbf{\emph{non-thre}}: This baseline does not consider collusion thresholds. It is a special case of Alg.~\ref{alg:DP_OPS_2} when $t = n-1$.
    \item \textbf{\emph{TMDP}}~\cite{nissim2011distributed}: This baseline does not consider personalized privacy budgets. Instead, each party's privacy budget is replaced with the most stringent one. This approach is also a special case of Alg.~\ref{alg:DP_OPS_2}.
    \item \textbf{\emph{MIN}}: This is a centralized baseline where a trusted third party gathers data from all parties, computes the query function with additive Gaussian noise, and then releases the results to the active parties.
    \item \textbf{\emph{Sample}}~\cite{jorgensen2015conservative}: This state-of-the-art centralized PDP mechanism divides parties into stringent and non-stringent groups based on a predetermined budget $\epsilon^{(t)}$. The data for parties with stringent privacy budgets are ignored with high probability, leading to a decrease in the required noise level. Following~\cite{jorgensen2015conservative}, we take $\epsilon^{(t)} = \frac{1}{n} \Sigma_{i\in [n]} \epsilon_i$ to achieve good accuracy results in various tasks.
    \item \textbf{\emph{PLDP}}~\cite{kairouz2015secure}: We also include the PLDP mechanism as a baseline, which uses the randomized response for count and input perturbation for linear regression.
    \item
    \textbf{\emph{non-pri}}:
    For linear regression, we compare our mechanism with the non-private linear regression algorithm.
\end{itemize}

\textbf{Datasets}: We evaluate mechanisms for count on synthetic data. We sample a dataset with $n$ binary values, with a default value of $n=1000$.
The density parameter $\rho$, which controls the fraction of 1's in the dataset, is set to 0.15 by default.
For linear regression, we use a real-world revenue dataset~\cite{ipums} that contains 5 features: the number of children, gender, age, educational level, and annual income for prediction. This dataset comprises 2.5 million records. For training, we use a default value of $n = 5\times 10^4$ data points, which are normalized to the range of [-1, 1] for each attribute.
In each experiment, the data are distributed randomly among all parties.

\textbf{Privacy budgets and thresholds}: We adopt a methodology similar to~\cite{jorgensen2015conservative} to randomly assign parties into three groups based on their privacy consideration. The groups are 1) conservative, comprising parties with high privacy consideration, 2) moderate, comprising parties with medium consideration, and 3) liberal, comprising parties with low consideration. We denote the fractions of conservative and moderate parties as $f_C$ and $f_M$, respectively. The remaining fraction of liberal parties is represented by $f_L=1-f_C-f_M$. The default values for $f_C$ and $f_M$ are set to $0.54$ and $0.37$, respectively.
We set the privacy budget parameters $\delta_i = \frac{1}{10n}$ for $i\in [n]$, which is advised as a small multiple of $\frac{1}{n}$~\cite{abadi2016deep}. To select values for $\epsilon_i, i\in [n]$, we randomly choose values for parties in conservative and moderate groups from the intervals $[\epsilon_C, \epsilon_M]$ and $[\epsilon_M, \epsilon_L]$, respectively, while fixing $\epsilon_i = \epsilon_L$ for liberal parties. The values for $\epsilon_C$, $\epsilon_M$, and $\epsilon_L$ are $0.01$, $0.2$, and $1.0$, respectively.
In our experiments, we set the threshold parameter $t$ to $\lfloor 0.5 n \rfloor$ by default.

\begin{figure*}[h]
\centering

\includegraphics[width=0.75\linewidth]{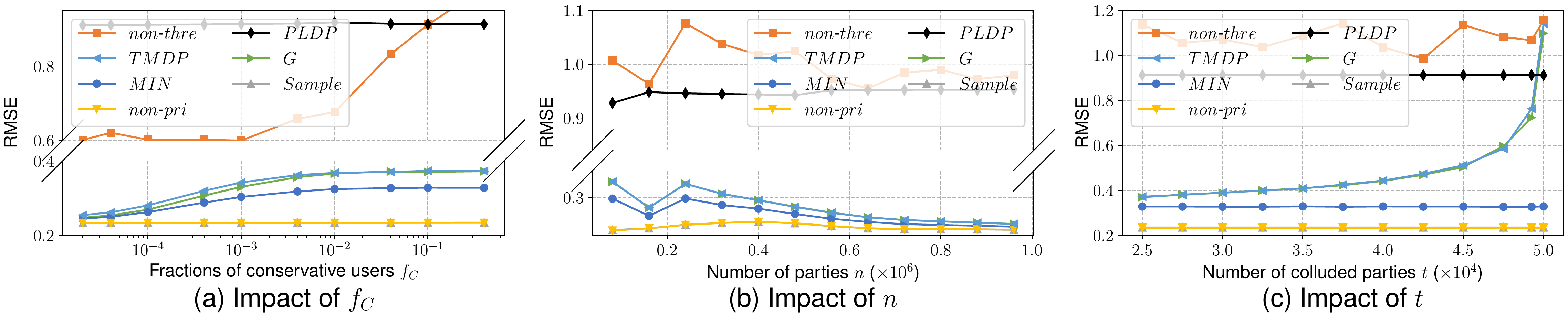}

\caption{Utility loss of different mechanisms for linear regression}\label{fig:linReg}
\vspace{-0.1in}
\end{figure*}

\textbf{Results for count}:
Fig.~\ref{fig:cnt} displays the utility of different mechanisms for count queries. To obtain these results, we randomly select ${U^+}$ from subsets of $[n]$. We repeat each mechanism for each configuration over 100 times and present the average results.
Overall, when $t = \lfloor 0.5 n \rfloor$, \emph{non-thre} performs significantly worse than other mechanisms. \emph{G}, \emph{TMDP}, and \emph{MIN} achieve similar utility in most cases. \emph{G} performs better than \emph{TMDP} when the fraction of conservative parties $f_c$ is relatively small, or when the threshold value $t$ is close to $n$.

1) \emph{Impact of data density}:
Fig.~\ref{fig:cnt}a shows that the utility loss of \emph{G}, \emph{TMDP}, \emph{MIN}, \emph{non-thre}, and \emph{PLDP} remains constant when data density $\rho$ varies. The utility of \emph{G} and \emph{TMDP} is similar to \emph{MIN}'s and much better than that of \emph{non-thre} and \emph{PLDP}. The utility loss of \emph{Sample} increases linearly with $\rho$. This can be explained by the fact that for smaller values of $\rho$, \emph{Sample} discards mostly 0's, resulting in a minor loss of accuracy. For larger $\rho$, however, \emph{Sample} ignores too many 1's, leading to a larger count error.

\hidemass{
2) \emph{Impact of privacy budgets}:
In Fig.~\ref{fig:cnt_b}, the utility loss of \emph{Sample} is not sensitive to $\epsilon_C$ since the additive noise of \emph{Sample} is determined by $\epsilon^{(t)}$, which is much larger than $\epsilon_C$~\cite{jorgensen2015conservative}. The utility loss of \emph{G}, \emph{TMDP}, \emph{MIN}, \emph{non-thre}, and \emph{PLDP} is large when $\epsilon_C$ is close to 0 and decreases rapidly as $\epsilon_C$ increases. When $\epsilon_C > 0.05$, \emph{G}, \emph{TMDP}, and \emph{MIN} perform the best among the candidate mechanisms and achieve a small and stable utility loss over varying values of $\epsilon_C$.
}

2) \emph{Impact of the fraction of conservative users}:
Fig.~\ref{fig:cnt}b demonstrates that \emph{G} outperforms \emph{TMDP} when $f_C \leq 0.04$. The utility of \emph{PLDP} deteriorates sharply when $f_C \geq 0.1$. The centralized mechanism \emph{MIN}'s utility is similar to that of \emph{G} and \emph{TMDP}.
\emph{Sample} achieves the best utility since
the additive noise is determined by
$\epsilon^{(t)}$, which is much larger than $\epsilon_C$.

3) \emph{Impact of the number of parties}:
Regarding the impact of $n$ shown in Fig.~\ref{fig:cnt}c, \emph{MIN}, \emph{G}, and \emph{TMDP} attain the lowest utility loss for $n\geq 5000$. Other mechanisms experience increasing utility loss as $n$ increases. This is because, for \emph{Sample}, the discarded data increase linearly with $n$, while for \emph{non-thre} and \emph{PLDP}, the noise variance increases linearly with $n$.

4) \emph{Impact of the collusion threshold}:
Fig.~\ref{fig:cnt}d examines the impact of $t$ and shows that except for \emph{G} and \emph{TMDP}, other mechanisms are not sensitive to changes in $t$, as only \emph{G} and \emph{TMDP} consider the collusion thresholds. The utility loss of \emph{G} is similar to that of \emph{MIN} for $t<700$ and better than that of \emph{PLDP} when $t<900$. Comparing \emph{G} and \emph{TMDP}, they have similar utility for $t \leq 900$. When $t > 900$, \emph{G} achieves significantly better utility than \emph{TMDP}.

\textbf{Results for linear regression}: We employ the Functional mechanism proposed in~\cite{zhang2012functional} to perform linear regression, where we solve the weight of the linear regression loss function with its coefficients perturbed by the noises generated by our and the baseline Gaussian mechanisms.
For each experiment, we conduct 20 runs of five-fold cross-validation.

In this study, we also investigate the impact of $f_C$, $n$, and $t$ on utility loss for different mechanisms. The results are presented in Fig.~\ref{fig:linReg}. Our evaluation reveals that \emph{non-thre} and \emph{PLDP} exhibit the largest utility loss. \emph{Sample} achieves the closest utility to \emph{non-pri}, which can be attributed to the smaller additive noise variance in \emph{Sample} than the other three mechanisms. 
\emph{G} and \emph{TMDP} exhibit close performance to \emph{MIN} in most cases.
Additionally, \emph{G} features better utility than \emph{TMDP} when the fraction of conservative parties $f_c$ is small, or when the threshold value $t$ is close to $n$.

\hidemass{
1) \emph{Impact of privacy budgets}: Fig.~\ref{fig:linReg_a} indicates that \emph{Sample} exhibits the same utility as \emph{non-pri}. The utility loss of \emph{G}, \emph{TMDP}, and \emph{MIN} decreases sharply as $\epsilon_C$ increases and becomes the same as \emph{Sample} and \emph{non-pri} when $\epsilon_C \geq 0.08$. We also observe that \emph{PLDP} and \emph{non-thre} perform poorly over the parameter settings for $\epsilon$.
}

1) \emph{Impact of the fraction of conservative users}: Our results in Fig.~\ref{fig:linReg}a reveal that the utility loss of \emph{G}, \emph{non-thre}, \emph{TMDP}, and \emph{MIN} increases with $f_C$. Comparing \emph{G} and \emph{TMDP}, we observe that \emph{G} outperforms TMDP slightly for $f_C \leq 0.01$.

2) \emph{Impact of the number of parties}: We observe, in Fig.~\ref{fig:linReg}b, that the utility loss for \emph{G}, \emph{TMDP}, and \emph{MIN} roughly decreases as $n$ increases since the increase in accuracy due to more training data exceeds the error caused by additive noise.
However, the loss fluctuation of \emph{G}, \emph{TMDP}, and \emph{MIN} around $n = 1.6\times 10^5$ occurs because the accuracy increase due to more training data does not dominate the noise when $n$ is not sufficiently large. The utility of \emph{G}, \emph{TMDP}, and \emph{MIN} converges to that of \emph{non-pri} and \emph{Sample} when $n\geq 9\times 10^5$.

3) \emph{Impact of the collusion threshold}: Fig.~\ref{fig:linReg}c demonstrates the impact of $t$. When $t \leq 2500$, the utility loss of \emph{G} and \emph{TMDP} is similar to that of \emph{MIN}. As $t$ increases, the utility of \emph{G} and \emph{TMDP} improves significantly and eventually reaches the maximum value when $t$ approaches $n-1$. When $t$ is close to $5\times 10^4$, the utility of \emph{G} surpasses that of \emph{TMDP}.

\subsection{Scalability}\label{subsec:scala}

\begin{figure}[ht]
\centering
\includegraphics[width=\linewidth]{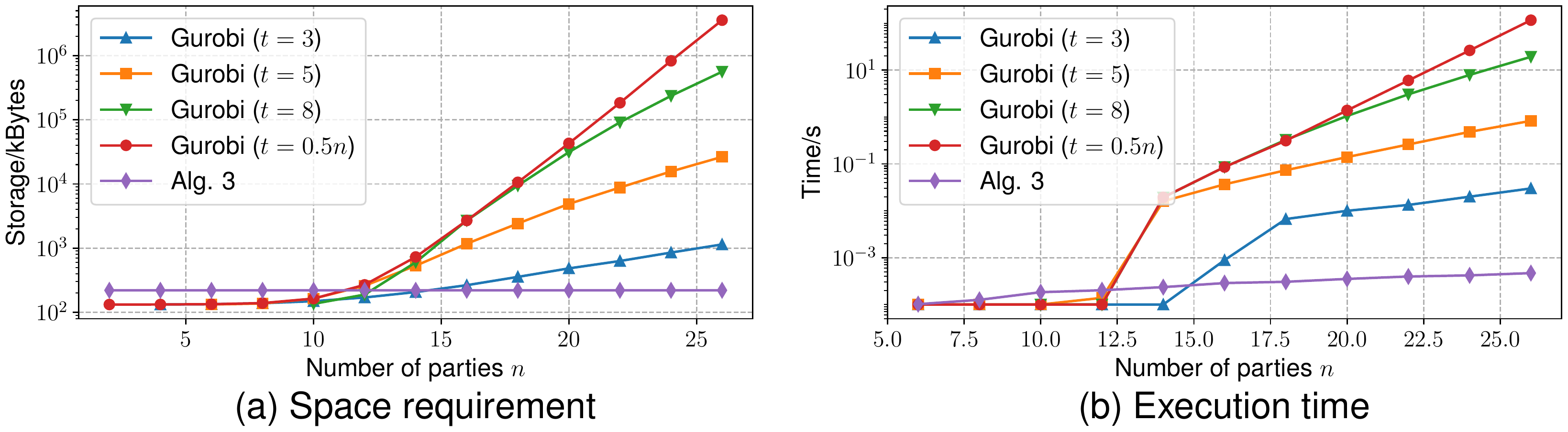}
\caption{Efficiency evaluation. The space requirement and execution time of Alg.~\ref{alg:DP_OPS_2} are independent of $t$.}\label{fig:eff}
\end{figure}

In this study, we evaluate the efficiency of Alg.~\ref{alg:DP_OPS_2} by comparing it with Gurobi~\cite{gurobi}, one of the most efficient LP solvers. Fig.~\ref{fig:eff}a presents a comparison of the space cost of Gurobi and our algorithm for varying values of $n$. We observe that the space requirement of Alg.~\ref{alg:DP_OPS_2}
is insensitive to the change of $n$ and $t$,
and is significantly smaller than that of Gurobi for $n \geq 15$. However, for Gurobi, the space complexity increases as $t$ increases for varying values of $n$. Specifically, when $t = 0.5n$, the space complexity of Gurobi increases exponentially with $n$.
Additionally, we analyze the execution time of Gurobi and Alg.~\ref{alg:DP_OPS_2} with varying values of $n$, as shown in Fig.~\ref{fig:eff}b. Consistent with the space requirement analysis in Table~\ref{tab:complexity}, the execution time of Gurobi increases as $t$ increases for varying $n$, while Alg.~\ref{alg:DP_OPS_2} exhibits constant execution time for different $t$. As for the impact of $n$, we note that when $n\leq 12$, the execution time for both Gurobi and Alg.~\ref{alg:DP_OPS_2} is less than 1 millisecond. However, for larger values of $n$, the execution time of Gurobi increases sharply (even exponentially for $t = 0.5n$), while the execution time of Alg.~\ref{alg:DP_OPS_2} increases slowly with $n$.
In summary, our results demonstrate that Alg.~\ref{alg:DP_OPS_2} outperforms Gurobi in terms of both space complexity and execution time for large values of $n$, especially when $t$ is a constant factor of $n$.


\section{Related Work}\label{sec:related}

\textbf{Secure multi-party computation (MPC)}.
MPC
is a cryptographic technique that enables multiple parties to collaboratively compute a function while ensuring that no collusion of $t$ parties can learn more than the outcome.
The extensive literature on MPC provides a variety of protocols for computing a broad range of functions~\cite{yao1986generate,ben1988completeness,goldreich2007foundations}, and researchers have focused on reducing the computation and communication costs of MPC protocols~\cite{ishai2003extending,demmler2015aby}. However, MPC has a drawback in that disclosing the exact function output may reveal a party's private
information,
especially when some parties collude.

\textbf{Differential privacy (DP)}.
DP
provides a methodology
to protect individual input entries by introducing sufficient randomness to the input (input perturbation) or the computation outcome (output perturbation)~\cite{dwork2014algorithmic}.
Mironov et al.~\cite{mironov2009computational} and Jorgensen et al.~\cite{jorgensen2015conservative} have respectively generalized DP to handle cases where the adversaries' computation power is bounded and users' privacy budgets are personalized. DP mechanisms provide a theoretical guarantee for protecting the private information contained in the output, albeit at the cost of sacrificing data utility due to the applied perturbation. To address this issue, researchers have developed a series of methods to find the optimal DP mechanism with respect to utility~\cite{geng2014optimal,geng2015optimal,balle2018improving}.

\textbf{Multi-party differential privacy}.
Multi-party differential privacy aims to
extend DP to
accommodate the settings of multi-party computations. Dwork et al.~\cite{dwork2006our} proposed a securely distributed noise generation protocol that enables the construction of multi-party DP mechanisms. Beimel et al.~\cite{nissim2011distributed} provided a formal definition for threshold multi-party differential privacy (TMDP) and presented two methods for constructing TMDP mechanisms: one by combining MPC and DP, and the other based on local DP (LDP) methods. Combining MPC and DP typically results in better utility, and many privacy-preserving computation systems with stringent utility requirements are constructed using this approach~\cite{goryczka2013secure,papadimitriou2017dstress,acar2017achieving}.
\comment{Recently, Tang et al.~\cite{TangCJLG22} proposed a multi-party personalized DP mechanism for marginal release by combining MPC with the Laplace mechanism satisfying DP.}
In contrast, the LDP methods are highly efficient at the expense of utility.
Murakami et al.~\cite{Murakami019} introduced the notion of
personalized LDP and optimized its utility.
Recently, researchers have proposed several personalized LDP mechanisms for machine learning~\cite{LiYCSL23} and statistics tasks~\cite{ShenXY21}.
Moreover, Cheu et al.~\cite{cheu2019distributed} designed multi-party DP mechanisms using
an anonymous channel called a shuffler, whose utility is strictly between LDP and centralized DP.

Despite the numerous contributions made in various aspects, there is a noticeable gap in designing multi-party DP mechanisms for the threshold model that allows for personalized privacy budgets while minimizing utility loss.

\section{Conclusions}
\label{sec:conc}
We present TPMDP, a novel general framework for achieving multi-party differential privacy. Our approach enables parties to privately compute a desired function in the presence of at most $t$ colluding semi-honest adversaries while outputting the results to a specified subset.
\comment{In this process, each party's personalized privacy requirement is met.}
We propose an easy-to-implement TPMDP mechanism that leverages the Gaussian mechanism. To minimize the overall variance of the Gaussian additive noise, we formulate the problem as an LP and provide an exact algorithm with linear complexity in the number of parties.
Our experiments demonstrate the effectiveness of our approach in terms of low utility loss, storage requirements, and running time.
{
\appendices
\hide{
\section{Notations}
    \renewcommand\arraystretch{1}
    \begin{table}[htbp]
        \begin{tabular}{>{\centering\arraybackslash} p{1.5cm} |  m{6.3cm} }
        \hline
        \multicolumn{2}{c}{MPC} \\
        \hline
        $U$        & Parties in an MPC protocol or TPMDP mechanism \\
        \hline
        $A$ & Adversary set \\
        \hline
        $\mathcal{A}$, $\mathcal{A}_t(\mathcal{A}_{t})$ & Adversary structure, $t$($t$)-threshold adversary structure \\
        \hline
        $\mathcal{V}_{A}^{\Pi}({\boldsymbol{x}})$ & Viewed values of parties in $A$ during an execution of protocol $\Pi$ with input ${\boldsymbol{x}}$ \\
        \hline
        \multicolumn{2}{c}{DP} \\
        \hline
        $(\epsilon, \delta)$ & Privacy budget in DP \\
        \hline
        $\Delta_2 f$ & $l_2$-sensitivity of function $f$ \\
        \hline
        $\Gamma$ & The tuple $(\epsilon, \delta, \Delta_2 f)$ \\
        \hline
        $\sigma_{\Gamma}$ & The lower bound of the standard deviation of Gaussian mechanism to satisfy $(\epsilon, \delta)$-DP, cf. Definition \ref{def:sigma_Gamma}\\
        \hline
        $\simeq$ & Neighboring relation, e.g. $x$ and $y$ satisfy $x\simeq y$ if $x$ and $y$ differ in one single element\\
        \hline
        \multicolumn{2}{c}{TPMDP} \\
        \hline
        $\mathcal{RV}_{A}^{\Pi}({\boldsymbol{x}})$ & Refined view for $A$ during execution of a TPMDP mechanism $\Pi$ with input ${\boldsymbol{x}}$, cf. Definition \ref{def:rView} \\
        \hline
        $\mathcal{E}$ & The tuple consisting of privacy budgets of parties in $U$ \\
        \hline
        $\perp$ & A special symbol representing ``no information'' \\
        \hline
        $\Delta_{2,j} F$ & $j$-th partial $l_2$-sensitivity of $F$, cf. Definition \ref{def:p_l2_sensitivity} \\
        \hline
        ${U^+}$ & Set of active parties receiving an output unequal to $\perp$ in a multi-party Gaussian mechanism\\
        \hline
        ${\mathcal{A}_{t}^+}$ & Active $t$-threshold adversary structure consisting of all adversary sets containing at least one active party\\
        \hline
        $\boldsymbol{\sigma}$ & The tuple consisting of the standard deviations of noises in a multi-party Gaussian mechanism \\
        \hline
        $v_{\boldsymbol{\sigma}}$ & The objective function in Eq.~\eqref{eq:Gauss_opt}, i.e. $v_{\boldsymbol{\sigma}} = \Sigma_{i\in [n]} \sigma_i^2$ \\
        \hline
        $\overset{j}{\simeq}$ & $j$-neighboring relation, cf. Definition \ref{def:A_neigh_in} \\
        \hline
        \end{tabular}
        \caption{Notation.}
        \label{tab:notation}
    \end{table}
}

\hide{
\section{Definitions of Indistinguishability}\label{app:prel}

\begin{definition}[Computational indistinguishability~\cite{goldreich2007foundations}]\label{def:poly_ind}
    Two ensembles $X$ and $Y$ are computationally indistinguishable, if for every family $\{C_n\}_{n\in \mathbb{N}}$ of polynomial-size circuits, every positive polynomial $p(\cdot)$, and all sufficiently long $w$'s,
    \begin{equation}\label{eq:poly_ind}
        \left| \Pr[C_{|w|}(X_{w}) = 1] - \Pr[C_{|w|}(Y_w) = 1] \right| < \frac{1}{p(|w|)}.
    \end{equation}
\end{definition}


\begin{definition}[Statistical Indistinguishability~\cite{goldreich2007foundations}]\label{def:s_ind}
    Two ensembles $X$ and $Y$ are statistically indistinguishable, if for every positive polynomial $p(\cdot)$, and all sufficiently long $w$'s,
    \begin{equation}\label{eq:s_ind}
        \Sigma_{\alpha\in \{0,1\}^*} \left| \Pr[X_w =\alpha] - \Pr[Y_w =\alpha] \right| < \frac{1}{p(|w|)},
    \end{equation}
    \sout{$X$ and $Y$ are perfectly indistinguishable if the LHS of Eq.~\eqref{eq:s_ind} equals to $0$.}
\end{definition}


}

\section{Proofs in Section \ref{subsec:privacy-opt-objective}}\label{app:prob}
    In this section, we present the proof for Theorem~\ref{thm:DP_multi_Gaussian_stat} at the end of this section.

    \begin{proof}[Proof of Theorem~\ref{thm:DP_multi_Gaussian_stat}]\label{pf_thm_DP_multi_Gaussian_perf}
        %
        For the statistical case,
        since $\pi$ is statistically $\tau$-private (cf. Definition~\ref{def:c_MPC}), there exists a probabilistic polynomial-time algorithm $\mathcal{S}$ such that for all $A\in \mathcal{A}_{\tau}$,
        \begin{equation}\label{pf:DP_multi_Gaussian_perf}
            \begin{aligned}
                \{\mathcal{S}(A,{\boldsymbol{x}}_A,  {\boldsymbol{X}}_A,\Pi_A({\boldsymbol{x}},{\boldsymbol{X}}))\}_{{\boldsymbol{x}}\in (\{0,1\}^{*})^n,{\boldsymbol{X}}\in supp(\mathcal{P})} \overset{s}\equiv \\
                \{\mathcal{V}_A^\Pi ({\boldsymbol{x}},{\boldsymbol{X}})\}_{{\boldsymbol{x}}\in (\{0,1\}^{*})^n,{\boldsymbol{X}}\in supp(\mathcal{P})},
            \end{aligned}
        \end{equation}
        where ${\boldsymbol{x}}_A\triangleq \{x_i\}_{i\in A}$, ${\boldsymbol{X}}_A\triangleq \{X_i\}_{i\in A}$, and $\Pi_A({\boldsymbol{x}},{\boldsymbol{X}}) \triangleq \{\Pi_i({\boldsymbol{x}},{\boldsymbol{X}})\}_{i\in A}$.
        In Theorem~\ref{thm:DP_multi_Gaussian_stat}, we assume that $t \leq \tau$, which implies that $\mathcal{A}_t \subset \mathcal{A}_{\tau}$. Consequently,
        Eq.~\eqref{pf:DP_multi_Gaussian_perf} also holds for all $A\in \mathcal{A}_t$. Thus $(A,{\boldsymbol{x}}_A, {\boldsymbol{X}}_A,\Pi_A({\boldsymbol{x}},{\boldsymbol{X}}))$ is a refined view for $A\in \mathcal{A}_t$.

        Recall that
        $\Pi_j({\boldsymbol{x}},{\boldsymbol{X}})=F({\boldsymbol{x}})+\Sigma_{i=1}^n X_i$ for $j\in {U^+}$ and $\Pi_j({\boldsymbol{x}},{\boldsymbol{X}}) =\perp$ for $j\in {U^-}$.
        If $A\in {\mathcal{A}_{t}^+}$,
        it holds that $(A,{\boldsymbol{x}}_A,{\boldsymbol{X}}_A, F({\boldsymbol{x}})+\Sigma_{i\in \bar{A}} X_i)$ is a refined view for $A$, where $\Sigma_{i\in \bar{A}} X_i \sim \mathcal{N}(0, \Sigma_{i\in \bar{A}} \sigma_{i}^2)$.
        \comment{By Eq.~\eqref{eq:DP_multi_Gaussian_stat}, it holds that $\Sigma_{i\in \bar{A}} \sigma_{i}^2 \geq \sigma_{\Gamma_j}^2$ for all $j\in \Bar{A}$.
        As $\sigma_{\Gamma_j}^2$ denotes the sufficient noise variance for party $j$'s DP requirement, Eq.~\eqref{eq:tMDP} {holds} for all $j\in \bar{A}$.}
        On the other hand, if $A\notin {\mathcal{A}_{t}^+}$, then for all $j\in A$, $\Pi_j({\boldsymbol{x}},{\boldsymbol{X}})=\perp$. In this case, $(A,{\boldsymbol{x}}_A,{\boldsymbol{X}}_A, \perp)$ is a refined view.
        Thus Eq.~\eqref{eq:tMDP} {holds} for all $j\in \bar{A}$ as $(A,{\boldsymbol{x}}_A,{\boldsymbol{X}}_A, \perp) = (A,{\boldsymbol{x}}_A',{\boldsymbol{X}}_A, \perp)$ for $j$-neighboring ${\boldsymbol{x}}, {\boldsymbol{x}}'$.
        Overall, $\Pi$ satisfies $(t,n,\mathcal{E})$-TPMDP.
        %
        The proof for the computationally private case is similar.
    \end{proof}

\section{Proofs in Section~\ref{subsec:utility-max-alg}}\label{app:res-pf}
    \subsection{Proofs for Case 1: ${U^+} = [n]$}\label{case1_proof}
    \begin{lemma}\label{lemma:DP_sort}
        For Case 1, if $\boldsymbol{\sigma}$ is feasible for Eq.~\eqref{eq:Gauss_opt}, then $\hat{\boldsymbol{\sigma}} \triangleq \{\hat{\sigma}_i\}_{i\in [n]}$ is feasible, where $\hat{\boldsymbol{\sigma}}$ is a permutation of $\boldsymbol{\sigma}$ which satisfies $\hat{\sigma}_i \geq \hat{\sigma}_j$ when {$\sigma_{\Gamma_i}$} $\geq$ {$\sigma_{\Gamma_j}$}.
    \end{lemma}
    \begin{proof}
        Without loss of generality, assume that {$\sigma_{\Gamma_i} \geq \sigma_{\Gamma_j}$} for $i < j$.
        Suppose, by contradiction, that $\hat{\boldsymbol{\sigma}}$ is not feasible for Eq.~\eqref{eq:Gauss_opt}, which implies that there exists $k\in [n]$ such that the constraint in Eq.~\eqref{eq:Gauss_opt} does not hold.
        For ${U^+} = [n]$, we have ${\mathcal{A}_{t}^+} = \mathcal{A}_t$. As $\hat{\boldsymbol{\sigma}}$ is a permutation of $\boldsymbol{\sigma}$, it holds that $\{\min_{A_j} \Sigma_{i\in \bar{A}_j} \hat{\sigma}_i^2\}_{j=1}^n = \{\min_{A_j} \Sigma_{i\in \bar{A}_j} \sigma_i^2\}_{j=1}^n$, denoted by $E$, where $A_j$ is as defined in Eq.~\ref{eq:Gauss_opt}.
        Because $\boldsymbol{\sigma}$ is feasible for Eq.~\eqref{eq:Gauss_opt}, it holds that $\min_{A_j} \Sigma_{i\in \bar{A}_j} \sigma_i^2 \geq \sigma_{\Gamma_j}^2$ for $j\in [n]$. It follows that $\min_{A_j} \Sigma_{i\in \bar{A}_j} \sigma_i^2 \geq$ {$ \sigma_{\Gamma_k}^2$} $ > \min_{A_k} \Sigma_{i\in \bar{A}_k} \hat{\sigma}_i^2$ for all $j\in [k]$.
        By $\hat{\sigma}_i \geq \hat{\sigma}_j$ for $i<j$, $\min_{A_k} \Sigma_{i\in \bar{A}_k} \hat{\sigma}_i^2$ is the $k$-largest term in $E$; a contraction.
    \end{proof}

    
    By Lemma \ref{lemma:DP_sort}, we can restrict attention to those $\boldsymbol{\sigma}$'s satisfying the following convention as any feasible $\boldsymbol{\sigma}$ can be converted into a feasible $\hat{\boldsymbol{\sigma}}$ satisfying this convention.

    \begin{convention}\label{ass:Theta_sort_case1}
        $\boldsymbol{\sigma}$ satisfies that $\sigma_i \geq \sigma_j$ when {$\sigma_{\Gamma_i}\geq \sigma_{\Gamma_j}$}.
    \end{convention}
    
    For simplifying notation, without loss in generality, we assume that $\sigma_{\Gamma_i}\ge \sigma_{\Gamma_j}$, $\forall i<j$ in the following lemmas in this subsection, \comment{as formalized in Convention~\ref{ass:sigma_Theta_sort_case1}.}
    
    \begin{convention}\label{ass:sigma_Theta_sort_case1}
        $\{\sigma_{\Gamma_j}\}_{j\in [n]}$ satisfies $\sigma_{\Gamma_i}\ge \sigma_{\Gamma_j}$, $\forall i<j$.
    \end{convention}
    
    \begin{lemma}\label{lemma:sort_iff_cond}
        For Case 1, let Conventions~\ref{ass:Theta_sort_case1} and \ref{ass:sigma_Theta_sort_case1} hold. Then $\boldsymbol{\sigma}$ is feasible for Eq.~\eqref{eq:Gauss_opt} if {and only if }for $j\leq t+1$,
        \begin{equation}\label{eq:sort_iff_cond}
            \sigma_j^2+\Sigma_{i=t+2}^n \sigma_i^2 \geq \sigma_{\Gamma_j}^2.
        \end{equation}
    \end{lemma}

    \begin{proof}
        ``$\Rightarrow$'' {It follows} from
        Eq.~\eqref{eq:Gauss_opt} and Conventions \ref{ass:Theta_sort_case1} and \ref{ass:sigma_Theta_sort_case1}.

        ``$\Leftarrow$''
        By Conventions~\ref{ass:Theta_sort_case1} and \ref{ass:sigma_Theta_sort_case1}, it holds that $\min_{A_j} \Sigma_{i\in \bar{A}_j} \sigma_i^2 = \sigma_j^2+\Sigma_{i=t+2}^n \sigma_i^2 \geq \sigma_{\Gamma_j}^2$ for $j\leq t+1$. For $j \geq t+2$, we have $\min_{A_j} \Sigma_{i\in \bar{A}_j} \sigma_i^2 = \Sigma_{i=t+1}^n \sigma_i^2 \geq \sigma_{\Gamma_j}^2$. Thus,
        $\boldsymbol{\sigma}$ is feasible for Eq.~\eqref{eq:Gauss_opt}.
    \end{proof}

    \begin{lemma}\label{lemma:3cond}
        For Case 1, let Conventions~\ref{ass:Theta_sort_case1} and \ref{ass:sigma_Theta_sort_case1} hold. If $\boldsymbol{\sigma}$ is optimal for Eq.~\eqref{eq:Gauss_opt}, then
        \begin{enumerate}
            \item $\sigma_{t+1}^2=\sigma_{t+2}^2=\hdots =\sigma_n^2$;
            \item for $j\leq t$, if $\sigma_j^2 > \sigma_{j+1}^2$, then $\sigma_j^2+\Sigma_{i=t+2}^n \sigma_i^2 = $ {$\sigma_{\Gamma_j}^2$};
            \item $\sigma_1^2+\Sigma_{i=t+2}^n \sigma_i^2 = $ {$\sigma_{\Gamma_1}^2$}.
        \end{enumerate}
    \end{lemma}
    \begin{proof}[Sketch of proof]
        1) Assume, by contradiction, that condition 1) does not hold. Construct $\hat{\boldsymbol{\sigma}}$ where $\hat{\sigma}_{t+2}^2=\frac{1}{n-t-1} \Sigma_{i=t+2}^n \sigma_i^2 + \gamma$, $\hat{\sigma}_i^2=\frac{1}{n-t-1} \Sigma_{i=t+2}^n \sigma_i^2$ for $i \geq t+3$ and $\hat{\sigma}_i^2=\sigma_i^2-\gamma$ for $i\leq t+1$, for some $\gamma$ sufficiently small. It holds that $\hat{\boldsymbol{\sigma}}$ is feasible and $v_{\hat{\boldsymbol{\sigma}}} < v_{\boldsymbol{\sigma}}$; a contradiction.

        2) Assume, by contradiction, that condition 2) does not hold for $\boldsymbol{\sigma}$ when $j = k$. Then, construct $\hat{\boldsymbol{\sigma}}$ as $\hat{\sigma}_k^2 = \sigma_k^2 - \gamma$ and $\hat{\sigma}_j^2 = \sigma_j^2$ for $j \neq k$, for $\gamma$ sufficiently small. It holds that $\hat{\boldsymbol{\sigma}}$ is feasible and $v_{\hat{\boldsymbol{\sigma}}} < v_{\boldsymbol{\sigma}}$; a contradiction.

        3) If there exists $j\leq n-1$ such that $\sigma_j^2 > \sigma_{j+1}^2$, then condition 3) holds from condition 1) and condition 2). {Otherwise}, suppose, by contradiction, that condition 3) does not hold. It holds that $\hat{\boldsymbol{\sigma}}$ is feasible and $v_{\hat{\boldsymbol{\sigma}}} < v_{\boldsymbol{\sigma}}$ where $\hat{\sigma}_i^2 = \frac{1}{n-t} \sigma_{\Gamma_1}^2$ for $i\in [n]$; a contradiction.
    \end{proof}

    \begin{lemma}\label{lemma:3cond_cor}
        For Case 1, let Conventions~\ref{ass:Theta_sort_case1} and \ref{ass:sigma_Theta_sort_case1} hold. There always exists an optimal $\boldsymbol{\sigma}$ for Eq.~\eqref{eq:Gauss_opt} such that there exists $\xi\leq \min{(\lfloor \frac{2n-t}{n-t} \rfloor, t+1)}$, $\sigma_{\xi}^2 = \sigma_{\xi+1}^2 = \hdots = \sigma_{n}^2$ and $\sigma_j^2+\Sigma_{i=t+2}^n \sigma_i^2 = \sigma_{\Gamma_j}^2$ for $j\leq \xi$.
    \end{lemma}

    \begin{proof}[Sketch of proof]
        If the optimal $\boldsymbol{\sigma}$ satisfyies $\sigma_1^2 = \hdots = \sigma_n^2$, we simply take $\xi = 1$ in Lemma~\ref{lemma:3cond_cor}.
        Lemma \ref{lemma:3cond_cor} then holds from condition 3) in Lemma \ref{lemma:3cond}.
        
        If the optimal $\boldsymbol{\sigma}$ does not satisfy $\sigma_1^2 = \hdots = \sigma_n^2$, we choose $\xi = \arg \max_j \sigma_{j-1}^2 > \sigma_{j}^2$.
        By condition 1) in Lemma \ref{lemma:3cond}, we have $\xi\leq t + 1$.
        It remains to prove $\sigma_j^2+\Sigma_{i=t+2}^n \sigma_i^2 = \sigma_{\Gamma_j}^2$ for $j\leq \xi$ and $\xi \leq \lfloor \frac{2n-t}{n-t} \rfloor$.
        
        For $j\leq \xi-1$, by Conventions~\ref{ass:Theta_sort_case1} and \ref{ass:sigma_Theta_sort_case1} and condition 2) in Lemma \ref{lemma:3cond}, it holds that $\sigma_j^2+\Sigma_{i=t+2}^n \sigma_i^2 = \sigma_{\Gamma_j}^2$.
        For $j = \xi$, if $\boldsymbol{\sigma}$ does not satisfy $\sigma_{\xi}^2+\Sigma_{i=t+2}^n \sigma_i^2 = \sigma_{\Gamma_{\xi}}^2$, construct $\hat{\boldsymbol{\sigma}}$ where $\hat{\sigma}_j^2 = \sigma_j^2 - \gamma$ for $j\geq \xi$ and $\hat{\sigma}_j^2 = \sigma_j^2 + (n-t-1)\gamma$ for $j\leq \xi-1$ with $\gamma = \sigma_{\xi}^2-\frac{1}{n-t} \sigma_{\Gamma_{\xi}}^2$. It holds that $\hat{\boldsymbol{\sigma}}$ is feasible and $v_{\hat{\boldsymbol{\sigma}}}\leq v_{\boldsymbol{\sigma}}$, which implies that there exists an optimal $\hat{\boldsymbol{\sigma}}$ for Eq.~\eqref{eq:Gauss_opt} satisfying Lemma \ref{lemma:3cond_cor}.
        
        To prove $\xi \leq \lfloor \frac{2n-t}{n-t} \rfloor$, {suppose}, by contradiction, that $\xi > \lfloor \frac{2n-t}{n-t} \rfloor$. Construct $\hat{\boldsymbol{\sigma}}$ where $\hat{\sigma}_j^2 = \sigma_j^2 + \gamma$ for $j\geq \xi$ and $\hat{\sigma}_j^2 = \sigma_j^2 - (n-t-1)\gamma$ for $j\leq \xi-1$ with sufficiently small $\gamma$. {It holds that} $v_{\hat{\boldsymbol{\sigma}}} < v_{\boldsymbol{\sigma}}$, which yields a contradiction.
    \end{proof}

    \begin{proof}[Proof of Theorem \ref{thm:case_trivial}]
    \label{pf_lemma_exact_param}
        By Lemma \ref{lemma:3cond_cor}, the optimal $v_{\boldsymbol{\sigma}}$ satisfying the requirement in Lemma \ref{lemma:3cond_cor} equals $\Sigma_{i=1}^{\xi-1} \sigma_{\Gamma_j}^2 + (\frac{2n-t}{n-t}-\xi) \sigma_{\Gamma_{\xi}}^2$ (denoted by $\mathcal{L}_\xi$) for some $\xi\leq \min{(\lfloor \frac{2n-t}{n-t} \rfloor, t+1)}$. It holds that $\mathcal{L}_{\xi-1}-\mathcal{L}_{\xi} \geq 0$ for $\xi\leq \min{(\lfloor \frac{2n-t}{n-t} \rfloor, t+1)}$, which implies that $\boldsymbol{\sigma}$ in Theorem \ref{thm:case_trivial} is optimal.
    \end{proof}

    \subsection{Proofs in Case 2: ${U^+}\subset [n]$}

    For Case 2, the next lemma shows that a specific permutation of a feasible $\boldsymbol{\sigma}$ is also feasible. We omit the proof for length considerations. The proof is similar to that of Lemma \ref{lemma:DP_sort}, the main difference being that we need to evaluate the permutations of ${\boldsymbol{\sigma}^+}$ and ${\boldsymbol{\sigma}^-}$ separately.

    \begin{lemma}\label{lemma:DP_sort_active}
        If $\boldsymbol{\sigma} = \boldsymbol{\sigma}^+ \cup \boldsymbol{\sigma}^-$ is feasible for Eq.~\eqref{eq:Gauss_opt}, then ${\boldsymbol{\hat{\sigma}}} = \boldsymbol{\hat{\sigma}}^+ \cup \boldsymbol{\hat{\sigma}}^-$ is feasible for Eq.~\eqref{eq:Gauss_opt}, where ${{\boldsymbol{\hat{\sigma}}^+}}$ is a permutation of ${\boldsymbol{\sigma}^+}$ with ${\hat{\sigma}_{i^+}} \geq {\hat{\sigma}_{j^+}}$ if $\sigma_{\Gamma_{i^+}} \geq \sigma_{\Gamma_{j^+}}$,
        and ${{\boldsymbol{\hat{\sigma}}^-}}$ is a permutation of ${\boldsymbol{\sigma}^-}$ with ${\hat{\sigma}_{i^-}} \geq {\hat{\sigma}_{j^-}}$ if $\sigma_{\Gamma_{i^-}} \geq \sigma_{\Gamma_{j^-}}$.
    \end{lemma}

    By Lemma \ref{lemma:DP_sort_active}, we can conclude that we can restrict attention to $\boldsymbol{\sigma}$'s satisfying the following convention.

    \begin{convention}\label{ass:Theta_sort_case2}
        $\boldsymbol{\sigma}$ satisfies that $\sigma_{i^+}^2 \geq \sigma_{j^+}^2$ when $\sigma_{\Gamma_{i^+}} \geq \sigma_{\Gamma_{j^+}}$ and $\sigma_{{i^-}}^2 \geq \sigma_{{j^-}}^2$ when $\sigma_{\Gamma_{i^-}} \geq \sigma_{\Gamma_{j^-}}$.
    \end{convention}
    
    Similarly, as in Case $1$, without loss in generality, we assume that $\sigma_{\Gamma_{i^+}} \geq \sigma_{\Gamma_{j^+}}$, $\forall i<j$ and $\sigma_{\Gamma_{i^-}} \geq \sigma_{\Gamma_{j^-}}$, $\forall i<j$ in the following lemmas, as formalized in Convention~\ref{ass:sigma_Theta_sort_case2}.
    
    \begin{convention}\label{ass:sigma_Theta_sort_case2}
        $\{\sigma_{\Gamma_j}\}_{j\in [n]}$ satisfies $\sigma_{\Gamma_{i^+}} \geq \sigma_{\Gamma_{j^+}}$, $\forall i<j$ and $\sigma_{\Gamma_{i^-}} \geq \sigma_{\Gamma_{j^-}}$, $\forall i<j$.
    \end{convention}
    
    \begin{lemma}\label{lemma:2*_trivial}
        Let Conventions \ref{ass:Theta_sort_case2} and \ref{ass:sigma_Theta_sort_case2} hold and $2 \leq \left| {U^+} \right| \leq n-t$. If $\boldsymbol{\sigma}$ is optimal for Eq.~\eqref{eq:Gauss_opt} and $\sigma_{2^+}^2 \geq \sigma_{t^-}^2$, then
        $v_{\hat{\boldsymbol{\sigma}}} \leq v_{\boldsymbol{\sigma}}$ if $\hat{\boldsymbol{\sigma}}$ is optimal for Eq.~\eqref{eq:Gauss_opt} when replacing $\mathcal{A}_{t}^+$ with $\mathcal{A}_{t}$.
    \end{lemma}

    \begin{proof}[Sketch of proof]
        By Eq.~\eqref{eq:Gauss_opt},
        a feasible $\boldsymbol{\sigma}$ satisfying $\sigma_{2^+}^2 \geq \sigma_{t^-}^2$ remains feasible when replacing ${\mathcal{A}_{t}^+}$ by ${\mathcal{A}_{t}}$.
        Thus,
        $v_{\hat{\boldsymbol{\sigma}}} \leq v_{\boldsymbol{\sigma}}$ if $\hat{\boldsymbol{\sigma}}$ is optimal for Eq.~\eqref{eq:Gauss_opt} when replacing $\mathcal{A}_{t}^+$ with $\mathcal{A}_{t}$.
    \end{proof}

    By Lemma \ref{lemma:2*_trivial}, if $\boldsymbol{\sigma}$ satisfying $\sigma_{2^+}^2\geq \sigma_{t^-}^2$ under Convention \ref{ass:Theta_sort_case2} is optimal for Eq.~\eqref{eq:Gauss_opt},
    any $\hat{\boldsymbol{\sigma}}$ derived from Alg. \ref{alg:DP_OPS_trivial} is also optimal.
    Thus, we restrict attention to a special class of $\boldsymbol{\sigma}$'s specified in the following convention.

    \begin{convention}\label{ass:2*_tau_cmp}
        Let Convention~\ref{ass:sigma_Theta_sort_case2} hold
        and $2 \leq \left| {U^+} \right| \leq n-t$. $\boldsymbol{\sigma}$ satisfies $\sigma_{2^+}^2\leq \sigma_{t^-}^2$.
    \end{convention}

    \begin{lemma}\label{lemma:sort_iff_cond_active}
        Let Convention \ref{ass:2*_tau_cmp} hold. $\boldsymbol{\sigma}$ is feasible for Eq.~\eqref{eq:Gauss_opt}~iff
        \begin{equation}\label{eq:sort_iff_cond_active}
        \left\{
        \begin{aligned}
            &\sigma_j^2+\Sigma_{i\in E} \sigma_i^2 \geq \sigma_{\Gamma_j}^2,\, j\in \{1^-,\hdots,t^-\} \\
            &\sigma_{t^-}^2+\Sigma_{i\in E} \sigma_i^2 \geq \sigma_{\Gamma_{2^+}}^2 \\
            &\sigma_{{1^+}}^2+\Sigma_{i\in E} \sigma_i^2-\sigma_{2^+}^2+\sigma_{t^-}^2 \geq \sigma_{\Gamma_{{1^+}}}^2,
        \end{aligned}
        \right.
        \end{equation}
        where $E=\{2^+,\hdots,(\eta+1)^+, (t+1)^-, \hdots, (n-\eta-1)^-\}$.
    \end{lemma}
    \begin{proof}[Sketch of proof]
        Eq.~\eqref{eq:sort_iff_cond_active} is equivalent to the constraints in Eq.~\eqref{eq:Gauss_opt} under Convention \ref{ass:2*_tau_cmp}.
    \end{proof}

    \begin{lemma}\label{lemma:DP_2cond_active}

        Let Convention \ref{ass:2*_tau_cmp} hold. Then $\boldsymbol{\sigma}$ is feasible for Eq.~\eqref{eq:Gauss_opt} iff $\boldsymbol{\sigma}$ is feasible when replacing $\{\sigma_{\Gamma_i}\}_{i\in [n]}$ by $\{\tilde{\sigma}_{\Gamma_i}\}_{i\in [n]}$ where
        \begin{equation}\label{eq:sort_iff_cond_active4}
            \left\{
            \begin{aligned}
                &\tilde{\sigma}_{{\Gamma}_{i}} = \max\{ \sigma_{\Gamma_{i}}, \sigma_{\Gamma_{2^+}}\},\, i\in \{1^-, 2^-, \hdots, t^-\} \\
                &\tilde{\sigma}_{{\Gamma}_{j}}=\sigma_{\Gamma_j},\, \text{otherwise}.
            \end{aligned}
            \right.
        \end{equation}
    \end{lemma}
    \begin{proof}
        It follows from Lemma \ref{lemma:sort_iff_cond_active}.
    \end{proof}

    With Lemma \ref{lemma:DP_2cond_active}, we can restrict attention to a special class of $\{\sigma_{\Gamma_i}\}_{i\in [n]}$
    formalized in the following convention.

    \begin{convention}\label{ass:tau_2*_max_cond}
        Let Convention \ref{ass:2*_tau_cmp} hold, and $\sigma_{\Gamma_j}, j\in [n]$ satisfy
        $$
            \sigma_{\Gamma_i}\geq \sigma_{\Gamma_{2^+}},\, \text{for }i\in \{1^-,2^-\hdots,t^-\}.
        $$
    \end{convention}

    A corollary of Lemma \ref{lemma:sort_iff_cond_active} under Convention \ref{ass:tau_2*_max_cond} follows.

    \begin{corollary}\label{cor:sort_iff_cond_cor_active}
        Let Convention \ref{ass:tau_2*_max_cond} hold. $\boldsymbol{\sigma}$ is feasible for Eq.\eqref{eq:Gauss_opt}~iff
        \begin{equation}\label{eq:sort_iff_cond_cor_active}
        \left\{
        \begin{aligned}
            &\sigma_j^2+\Sigma_{i\in E} \sigma_i^2 \geq \sigma_{\Gamma_j}^2,\, \text{for }j\in \{1^-,2^-,\hdots,t^-\} \\
            &\sigma_{{1^+}}^2+\Sigma_{i\in E} \sigma_i^2-\sigma_{2^+}^2+\sigma_{t^-}^2 \geq \sigma_{\Gamma_{{1^+}}}^2,
        \end{aligned}
        \right.
        \end{equation}
        where $E$ is as defined in Lemma~\ref{lemma:sort_iff_cond_active}.
    \end{corollary}

    In Lemma~\ref{lemma:DP_3cond_active}, we present several necessary conditions for the optimal solutions of Eq.~\eqref{eq:Gauss_opt} under Convention \ref{ass:tau_2*_max_cond}. We omit the proof for length considerations. The proof is similar to that of Lemma \ref{lemma:3cond}, the main difference being that here we invoke the feasibility condition provided by Corollary \ref{cor:sort_iff_cond_cor_active}.

    \begin{lemma}\label{lemma:DP_3cond_active}
        Let Convention \ref{ass:tau_2*_max_cond} hold and $E$ be the set as defined in Lemma~\ref{lemma:sort_iff_cond_active}. If $\boldsymbol{\sigma}$ is optimal for Eq.~\eqref{eq:Gauss_opt}, then
        \begin{enumerate}
            \item if $t\geq 2$, $\sigma_{t^-}^2=\sigma_{(t+1)^-}^2=\hdots =\sigma_{(n-\eta-1)^-}^2$;
            \item $\sigma_{2^+}^2=\sigma_{3^+}^2=\hdots=\sigma_{(\eta+1)^+}^2$;
            \item for $j < t$, if $\sigma_{j^-}^2 > \sigma_{(j+1)^-}^2$, then $\sigma_{j^-}^2+\Sigma_{i\in E} \sigma_i^2  = \sigma_{\Gamma_{j^-}}^2$.
        \end{enumerate}
    \end{lemma}

    \begin{lemma}\label{lemma:DP_6cond_active}

        Let Convention \ref{ass:tau_2*_max_cond} hold and $\boldsymbol{\sigma}$ be optimal for Eq.~\eqref{eq:Gauss_opt}. Then,
        \begin{enumerate}
            \item If $n-t-t\eta\leq 0$, then $v_{\boldsymbol{\sigma}} \geq v_{\hat{\boldsymbol{\sigma}}}$, where $\hat{\boldsymbol{\sigma}}$ is optimal for Eq.~\eqref{eq:Gauss_opt} when replacing ${\mathcal{A}_{t}^+}$ by ${\mathcal{A}_{t}}$;
            \item If $n-t-t\eta> 0$, $\sigma_{2^+}^2=0$.
        \end{enumerate}
    \end{lemma}
    \begin{proof}
        For $n-t-t\eta \leq 0$, by Lemma \ref{lemma:2*_trivial}, it suffices to prove that there exists an optimal $\boldsymbol{\sigma}$ under Convention \ref{ass:tau_2*_max_cond} satisfying $\sigma_{2^+}^2 = \sigma_{t^-}^2$. If $\boldsymbol{\sigma}$ does not satisfy $\sigma_{2^+}^2 = \sigma_{t^-}^2$, construct $\hat{\boldsymbol{\sigma}}$ as $\hat{\sigma}_{j}^2 = \sigma_{j}^2 + \gamma$ for $j \in {U^+}$ and $\hat{\sigma}_j^2 = \sigma_j^2-\frac{\eta}{n-\eta-t}\gamma$ for $j \in {U^-}$ where $\gamma = \frac{2n-\eta-t}{n-\eta-t} (\sigma_{t^-}^2 - \sigma_{2^+}^2)$.
        Then, $\hat{\sigma}_{2^+}^2 = \hat{\sigma}_{t^-}^2$.
        Besides, $\hat{\boldsymbol{\sigma}}$ is feasible for Eq.~\eqref{eq:Gauss_opt} and $v_{\hat{\boldsymbol{\sigma}}} \leq v_{{\boldsymbol{\sigma}}}$.

        For $n-t-t\eta > 0$, assume, by contradiction, that $\boldsymbol{\sigma}$ does not satisfy $\sigma_{2^+}^2 = 0$. Construct $\hat{\boldsymbol{\sigma}}$ as $\hat{\sigma}_{j}^2 = \sigma_{j}^2 - \gamma$ for $j \in {U^+}$ and $\hat{\sigma}_j^2 = \sigma_j^2+\frac{\eta}{n-\eta-t}\gamma$ for $j \in {U^-}$ with a sufficiently small $\gamma$. Consequently, $\hat{\boldsymbol{\sigma}}$ is feasible for Eq.~\eqref{eq:Gauss_opt} under Convention \ref{ass:tau_2*_max_cond} and $v_{\hat{\boldsymbol{\sigma}}} < v_{\boldsymbol{\sigma}}$, a contradiction.
    \end{proof}

    
    \begin{proof}[Proof of Lemma \ref{lemma:DP_active_equiv}]
        For Subcase~\ref{subcase_1}, using \emph{Pigeonhole Principle}, $A_j$ in Eq.~\eqref{eq:Gauss_opt} is the same for $j\in [n]$ when the active set is ${U^+}$ or $[n]$. Thus,
        the feasible solutions for Eq.~\eqref{eq:Gauss_opt} in Cases 1 and 2 are identical.
        For Subcase~\ref{subcase_3}, Lemma~\ref{lemma:DP_active_equiv} follows from Lemmas \ref{lemma:DP_6cond_active} and \ref{lemma:2*_trivial}.
    \end{proof}

    \begin{proof}[Proof of Lemma \ref{lemma:{U^+}_1_trivial}]
        First, we prove $\sigma_{{1^+}}^2 = 0$. Assume, by~contradiction, that $\sigma_{{1^+}}^2 > 0$. Construct $\hat{\boldsymbol{\sigma}}$ with $\hat{\sigma}_{{1^+}}^2 = 0$~and~$\hat{\sigma}_j^2 = \sigma_j^2$ for $j\neq {1^+}$.
        $\hat{\boldsymbol{\sigma}}$ is feasible and $v_{\hat{\boldsymbol{\sigma}}} < v_{\boldsymbol{\sigma}}$;
        a contradiction.

        For $t \geq 2$, as we have proved that $\sigma_{{1^+}}^2 = 0$, we only need to show that
        $\boldsymbol{\sigma}$ is optimal for Eq.~\eqref{eq:Gauss_opt} if $\boldsymbol{\sigma}^-$ is optimal for Eq.~\eqref{eq:Gauss_opt} with parameters $(t-1,n-1,{\mathcal{E}^-})$ and active set ${U^-}$.
        As the objective function in the latter optimization problem remains the same when $\sigma_{{1^+}}^2 = 0$, it is sufficient to prove that the feasible solutions for these two problems coincide. As ${1^+}$ is the only active party in $[n]$, $\{\bar{A} | A\in {\mathcal{A}_{t}^+}\}$ remains the same in these two problems.
        By Eq.~\eqref{eq:Gauss_opt}, the feasible solutions for these two problems are identical.

        For $t = 1$,
        $\boldsymbol{\sigma}$ is feasible for Eq.~\eqref{eq:Gauss_opt} iff $\Sigma_{i=1}^{n-1} \sigma_{i^-}^2 \geq \max_{j \in {U^-}} \sigma_{\Gamma_j}^2$. So $v_{\boldsymbol{\sigma}} = \Sigma_{i=1}^{n-1} \sigma_{i^-}^2 = \max_{j \in {U^-}} \sigma_{\Gamma_j}^2$ is optimal.
    \end{proof}
    


    For Subcase
    \ref{subcase_4_ext},
    we present three necessary conditions for $\boldsymbol{\sigma}$ to be optimal for Eq.~\eqref{eq:Gauss_opt} under Convention \ref{ass:tau_2*_max_cond} in Lemma \ref{lemma:DP_5cond_active}. We omit the details for length considerations. The proof for the first condition is similar to that for condition 3) in Lemma \ref{lemma:3cond}, and the proofs for the other two conditions are similar to that of Lemma \ref{lemma:3cond_cor}.

    \begin{lemma}\label{lemma:DP_5cond_active}
        For Subcase \ref{subcase_4_ext}, let Convention \ref{ass:tau_2*_max_cond} hold and $E$ be the set defined in Lemma~\ref{lemma:sort_iff_cond_active}. If $\boldsymbol{\sigma}$ is optimal for Eq.~\eqref{eq:Gauss_opt}, then
        \begin{enumerate}
            \item $\sigma_{1^-}^2+\Sigma_{i\in E} \sigma_i^2  = \sigma_{\Gamma_{1^-}}^2$;
            \item there exists $1\leq \xi \leq t$ such that $\sigma_{\xi^-}^2 = \sigma_{(\xi + 1)^-}^2=\hdots=\sigma_{(n-\eta-1)^-}^2$, and $ \sigma_{j^-}^2+\Sigma_{i\in E} \sigma_i^2 =\sigma_{\Gamma_{j^-}}^2, \forall j \leq \xi-1$;
            \item $\xi$ as defined in condition 2) satisfies $\xi=1$ if $\sigma_{{1^+}}^2>0$ and $\xi\leq 2$ if $\sigma_{{1^+}}^2=0$.
        \end{enumerate}
    \end{lemma}

    \begin{proof}[Proof of Lemma \ref{lemma:nontrivial_noise_active_ext} (sketch)]
        When $t = 1$,
        \comment{without loss in generality, assume that Convention~\ref{ass:sigma_Theta_sort_case2} holds for $\{\sigma_{\Gamma_i}\}_{i\in [n]}$.}
        As $\boldsymbol{\sigma}$ satisfies Convention \ref{ass:2*_tau_cmp}, it holds that $\boldsymbol{\sigma}$ is feasible for Eq.~\eqref{eq:Gauss_opt} by Eq.~\eqref{eq:sort_iff_cond_active}. Assume, by contradiction, that
        there exists a feasible $\hat{\boldsymbol{\sigma}}$ such that $v_{\hat{\boldsymbol{\sigma}}} < v_{{\boldsymbol{\sigma}}}$.
        Notice that $\boldsymbol{\sigma}$ satisfies $v_{\boldsymbol{\sigma}} = \max \{\sigma_{\Gamma_{{1^+}}}^2, \sigma_{\Gamma_{1^-}}^2, \sigma_{\Gamma_{2^+}}^2\}$ while it holds that $v_{\hat{\boldsymbol{\sigma}}} \geq \max_{i\in [n]} \sigma_{\Gamma_{i}}^2 \geq \max \{\sigma_{\Gamma_{{1^+}}}^2, \sigma_{\Gamma_{1^-}}^2, \sigma_{\Gamma_{2^+}}^2\}$. Thus, we have $v_{\hat{\boldsymbol{\sigma}}} \geq v_{\boldsymbol{\sigma}}$; a contradiction.

        When $t \geq 2$,
        for condition 1), assume, by contradiction, that there does not exist an optimal solution satisfying Convention \ref{ass:2*_tau_cmp}.
        By Lemma~\ref{lemma:2*_trivial}, this assumption implies that
        $v_{\boldsymbol{\sigma}} > v_{\hat{\boldsymbol{\sigma}}}$,
        where $\hat{\boldsymbol{\sigma}}$ is derived from Alg.~\ref{alg:DP_OPS_trivial} with input $(F, (t, n, \mathcal{E}), i)$ for $i\in [n]$.
        On the other hand, we can reason that $v_{\boldsymbol{\sigma}} \leq v_{\hat{\boldsymbol{\sigma}}}$
        via a simple algebraic computation, which yields a contradiction. Subsequently, by Lemma \ref{lemma:DP_2cond_active}, we can transform Eq.~\eqref{eq:Gauss_opt} 
        into an equivalent problem by replacing $\{\sigma_{\Gamma_i}\}_{i\in [n]}$ with $(\tilde{\sigma}_{\Gamma_i})_{i\in [n]}$ as defined in Lemma \ref{lemma:DP_2cond_active}.
        It holds that $\boldsymbol{\sigma}$ in condition 1) satisfies the necessary conditions in Lemma \ref{lemma:DP_5cond_active} with $\xi = 1$.
        We can further show that $\boldsymbol{\sigma}$ is optimal for $\xi = 1$.
        By condition 3) in Lemma~\ref{lemma:DP_5cond_active}, we derive that $\xi$ can only take the value $1$ or $2$.
        Lemma~\ref{lemma:DP_5cond_active} additionally implies
        that the optimization objective for $\xi=1$ is strictly smaller than for $\xi=2$.
        To conclude, $\boldsymbol{\sigma}$ is optimal in this case.
        For condition 2), the proof is similar.
    \end{proof}
    
    \begin{proof}[Proof of Theorem \ref{thm:case_general_rst}]
        It follows from Lemmas~\ref{lemma:DP_active_equiv}-\ref{lemma:nontrivial_noise_active_ext}.
    \end{proof}
    
    \hide{
    \begin{proof}[Proof of Theorem \ref{thm:case_general_rst} (sketch)]
        $\ \ $\\
        \textbf{Optimality}:
        \begin{itemize}
            \item Lines $3$--$12$ in Alg. \ref{alg:DP_OPS_2} generate the optimal $\sigma_i,\ i\in [n]$  following Lemma \ref{lemma:{U^+}_1_trivial};
            \item Lines $14$--$16$ follow Lemma \ref{lemma:DP_active_equiv} and Lemma \ref{lemma:DP_6cond_active_cor};
            \item Lines $18$--$30$ follow Lemma \ref{lemma:tau_1_case_active};
            \item Lines $32$--$48$ follow Lemma \ref{lemma:nontrivial_noise_active}.
        \end{itemize}
    
        \noindent \textbf{Complexity}:
        \begin{itemize}
            \item if $t = n-1$, the complexity is $\mathcal{O}(1)$ for Lines $3$--$12$ \& $14$--$16$;
            \item if $\left| {U^+} \right| =1$ and $i\in {U^+}$, the complexity is $\mathcal{O}(1)$ for Lines $4$--$5$;
            \item otherwise, Alg. \ref{alg:DP_OPS_2} needs to compute $\sigma_{\Gamma_j}$ for $j\in [n]$, find the largest and $2$-largest elements in $(\sigma_{\Gamma_j})_{j\in {U^+}}$ and $(\sigma_{\Gamma_j})_{j\in {U^-}}$ (i.e., Lines $9$\&$18$--$20$), or execute Alg.~\ref{alg:DP_OPS_trivial} (i.e., Lines $7$\&$15$), whose complexity is $\mathcal{O}(n)$.
        \end{itemize}%
    \end{proof}
    }

\hide{

\section{Proof in Section \ref{sec:perform}}\label{app:perform}
    \begin{proof}[Proof of Theorem \ref{thm:v_G_lo} (sketch)]
        For Subcase \ref{subcase_1}, as it degenerates to Case 1, we only need to consider ${U^+} = [n]$. Without loss of generality, assume that $\sigma_{\Gamma_i}\ge \sigma_{\Gamma_j}$, $\forall i<j$. 
        For Setting \ref{setting_1}, by Alg. \ref{alg:DP_OPS_trivial}, $v_{G} = \sigma_{\Gamma_1}^2 + \frac{t}{n-t} \sigma_{\Gamma_2}^2$ when $n>2t$. As it holds that $v_{MIN} = \sigma_{\Gamma_1}^2$, it follows that: $\frac{v_G}{v_{MIN}} \leq 1+\frac{t}{n-t} = \mathcal{O}(1).$
        For Setting \ref{setting_2}, when $n>\frac{1}{p(1-p)}$, it holds that $v_{G} = \Sigma_{i=1}^{\xi-1} \sigma_{\Gamma_i}^2 + (\frac{2n-t}{n-t} - \xi) \sigma_{\Gamma_{\xi}}^2$ where $\xi = \lfloor 2 + \frac{p}{1-p} \rfloor$, which implies $\frac{v_G}{v_{MIN}} = \mathcal{O}(1)$.
        The proofs for Subcases $\ref{subcase_2}-\ref{subcase_5}$ are similar.
    \end{proof}

    \begin{proof}[Proof of Theorem \ref{thm:v_up_O}]
        Similar to the proof for Theorem \ref{thm:v_G_lo}.
    \end{proof}
}

\section{Composition Properties of TPMDP}\label{app:comp}
      
    In this appendix, we present the composition properties of TPMDP.
    Formally, we consider $m$ TPMDP mechanisms, $\Pi^j$ for $j\in [m]$, where $\Pi^j$ satisfies $(t, n, \mathcal{E}^j)$-TPMDP.
    We denote the composition of $\{\Pi^j\}_{j\in [m]}$ as $\Pi^{1:m}$, i.e., $\Pi^{1:m} (\cdot) \triangleq (\Pi^1 (\cdot), \Pi^2 (\cdot), \hdots, \Pi^m (\cdot))$.
    
    \begin{theorem}[Composition for TPMDP] \label{thm:m_comp_thm}
        Let the mechanism $\Pi^j$ satisfy $(t, n, \mathcal{E}^j)$-TPMDP for $j\in [m]$, where $\mathcal{E}^j = \{(\epsilon_i^j, \delta_i^j)\}_{i\in [n]}$.
        If $m = poly (|x|)$ for some polynomial $poly (\cdot)$, then $\Pi^{1:m}$ satisfies
        statistical (computational) $(t,n,\mathcal{E}^{1:m})$-TPMDP where $\mathcal{E}^{1:m} = \{(\Sigma_{j=1}^m \epsilon_i^j, \Sigma_{j=1}^m \delta_i^j)\}_{i\in [n]}$.
    \end{theorem}
    
    \begin{proof}
        Let $\mathcal{RV}_A^{\Pi^j} (\boldsymbol{x}, \boldsymbol{X}^j)$ be the statistical (computational) refined view for $A\in \mathcal{A}_{t}$ in $\Pi^j$.
        That is, there exists a probabilistic polynomial-time algorithm $\mathcal{S}^j$ such that $\mathcal{S}^j (A, \mathcal{RV}_A^{\Pi^j} (\boldsymbol{x}, \boldsymbol{X}^j))$ is statistically (computationally) indistinguishable from $\mathcal{V}_{A}^{\Pi^j} (\boldsymbol{x}, \boldsymbol{X}^j)$ (cf. Definition~\ref{def:rView}).
        
        By the \emph{hybrid technique}~\cite{goldreich2007foundations},
        it holds that
        $\{\mathcal{S}^j (A, \mathcal{RV}_A^{\Pi^j} (\boldsymbol{x}, \boldsymbol{X}^j))\}_{j\in [m]}$ is statistically (computationally) indistinguishable from $\{\mathcal{V}_{A}^{\Pi^j} (\boldsymbol{x}, \boldsymbol{X}^j)\}_{j\in [m]}$ provided that $m=poly(|x|)$.
        Thus, $\mathcal{RV}_A^{\Pi^{1:m}} (\boldsymbol{x}, \boldsymbol{X}^{1:m}) \triangleq \{\mathcal{RV}_A^{\Pi^j} (\boldsymbol{x}, \boldsymbol{X}^j)\}_{j\in [m]}$ is a refined view for $\Pi^{1:m}$, where $\boldsymbol{X}^{1:m} = \{\boldsymbol{X}^{1}, \boldsymbol{X}^{2}, \hdots, \boldsymbol{X}^{m}\}$.
        
        By the fact that $\Pi^j$ satisfies $(t, n, \mathcal{E}^j)$-TPMDP,
        it holds that for all $j\in [m]$, all $A\in \mathcal{A}_{t}$, all $i\in \bar{A}$, all $i$-neighboring $\boldsymbol{x}, \boldsymbol{x}'$,
        and all sets $S^j$:
        $$
        \begin{aligned}
            \Pr[\mathcal{RV}_A^{\Pi^j}({\boldsymbol{x}},  {\boldsymbol{X}}^j) \in S^j] \leq
            e^{\epsilon_i^j} \Pr[\mathcal{RV}_A^{\Pi^j}({\boldsymbol{x}}',{\boldsymbol{X}}^j) \in S^j] + \delta_i^j.
        \end{aligned}
        $$
        By the composition theorem of DP (cf. Theorem 3.14 in~\cite{dwork2014algorithmic}), it follows that for all sets $S^{1:m}$:
        $$
        \begin{aligned}
            &\Pr[\mathcal{RV}_A^{\Pi^{1:m}}({\boldsymbol{x}},  {\boldsymbol{X}^{1:m}}) \in S^{1:m}] \leq \\
            &e^{\Sigma_{j=1}^m \epsilon_i^j} \Pr[\mathcal{RV}_A^{\Pi^{1:m}}({\boldsymbol{x}}',{\boldsymbol{X}}^{1:m}) \in S^{1:m}] + \Sigma_{j=1}^m \delta_i^j.
        \end{aligned}
        $$
        Thus, the composed mechanism $\Pi^{1:m}$ satisfies statistical (computational) $(t,n,\mathcal{E}^{1:m})$-TPMDP.
    \end{proof}
    
    For the $m$-fold composition of $(t, n, \mathcal{E})$-TPMDP mechanisms, we present an advanced composition theorem in Theorem~\ref{thm:m_A_comp}, which allows the privacy parameters to decay more slowly.
    We omit the proof,
    which is similar to that of Theorem \ref{thm:m_comp_thm}, the main difference being that in Theorem \ref{thm:m_A_comp}, the advanced composition theorem of DP (cf. Theorem 3.20 in \cite{dwork2014algorithmic}) is used.
    
    \begin{theorem}[Advanced Composition for TPMDP] \label{thm:m_A_comp}
        Let the mechanism $\Pi^j$ satisfy $(t, n, \mathcal{E})$-TPMDP for $j\in [m]$, where $\mathcal{E} = \{(\epsilon_i, \delta_i)\}_{i\in [n]}$.
        If $m = poly (|x|)$ for some polynomial $poly (\cdot)$, then for all $\delta_i^{1:m} > 0$ for $i\in [n]$,
        $\Pi^{1:m}$ satisfies
        statistical (computational) $(t,n,\mathcal{E}^{1:m})$-TPMDP where
        $\mathcal{E}^{1:m} = ((\epsilon_i, m{\delta}_i + \delta_i^{1:m}))_{i=1}^n$ and for $i\in [n]$,
        $$
        \epsilon_i^{1:m} = \sqrt{2m \ln{(1/\delta_i^{1:m})}} {\epsilon}_i + m {\epsilon}_i (e^{{\epsilon}_i}-1).
        $$
    \end{theorem}
    
    \hide{
    To prove the composition theorems for TPMDP, we first establish a lemma which gives a refined view for the composed mechanism $\Pi$ in Assumption~\ref{ass:comp_mech}.
    
    
    
    \begin{lemma}\label{lemma:RV_comb}
        Let Assumption \ref{ass:comp_mech} hold. For $A\in \mathcal{A}_t$, the composed refined view $(\mathcal{RV}_A^{\Pi^{(j)}}({\boldsymbol{x}},{\boldsymbol{X}}^{(j)}))_{j=1}^m$ is
        \hide{\begin{enumerate}
            \item a perfect refined view for $\Pi$ if $\mathcal{RV}_A^{\Pi^{(j)}}({\boldsymbol{x}},{\boldsymbol{X}}^{(j)})$ is a perfect refined view in $\Pi^{(j)}$ for $j\in \{1,\hdots,m\}$;
            \item a statistical (computational) refined view for $\Pi$ if $\mathcal{RV}_A^{\Pi^{(j)}}({\boldsymbol{x}},{\boldsymbol{X}}^{(j)})$ is a statistical (computational) refined view in $\Pi^{(j)}$ for $j\in \{1,\hdots,m\}$ and $m$ is polynomial related to $\left| {\boldsymbol{x}}\right|$.
        \end{enumerate}}
        a statistical (computational) refined view for $\Pi$ if $\mathcal{RV}_A^{\Pi^{(j)}}({\boldsymbol{x}},{\boldsymbol{X}}^{(j)})$ is a statistical (computational) refined view in $\Pi^{(j)}$ for $j\in \{1,\hdots,m\}$ and $m$ is polynomial related to $\left| {\boldsymbol{x}}\right|$.
    \end{lemma}
    
    Before presenting the proof, we first present an assumption.

    \begin{assumption}\label{ass:comp_en}
        Given are two collections of mutually independent ensembles $\{X^{(j)}\}_{j=1}^m$ and $\{X^{(j)}\}_{j=1}^m$, where 
        also $\{X^{(1)}, \hdots, X^{(i)}, Y^{(i+1)}, \hdots, Y^{(m)}\}$ are mutually independent collections, for each $i=1,\hdots,m-1$. Define the composed ensembles $\{(X_w^{(1)}, \hdots, X_w^{(m)})\}_{w\in E}$ and $\{(Y_w^{(1)}, \hdots, Y_w^{(m)})\}_{w\in E}$ to be $X$ and $Y$ respectively.
    \end{assumption}

    \begin{lemma}\label{lemma:comp_en}
        Let Assumption \ref{ass:comp_en} hold. Then
        \begin{enumerate}
            \item $X$ and $Y$ are perfectly indistinguishable if $X^{(i)}$ and $Y^{(i)}$ are perfectly indistinguishable for $i\in \{1,\hdots,m\}$;
            \item $X$ and $Y$ are computationally (statistically) indistinguishable if $X^{(i)}$ and $Y^{(i)}$ are computationally (statistically) indistinguishability for $i\in \{1,\hdots,m\}$ and $m$ is polynomial related to $\left| w \right|$.
        \end{enumerate}
    \end{lemma}
    \begin{proof}[Sketch of proof]
        For perfect and statistical indistinguishability, Lemma \ref{lemma:comp_en} can be proved by evaluating the LHS of Eq.~\eqref{eq:s_ind}. For computational indistinguishability, Lemma \ref{lemma:comp_en} can be proved using the \emph{hybrid technique}~\cite{goldreich2007foundations}.
    \end{proof}

    \begin{proof}[Proof of Lemma \ref{lemma:RV_comb}]
        For the perfect refined view, by Assumption \ref{ass:comp_mech}, $\{\mathcal{V}_A^\Pi ({\boldsymbol{x}},{\boldsymbol{X}}^{(j)})\}_{{\boldsymbol{x}}\in (\{0,1\}^{*})^n,{\boldsymbol{X}}^{(j)}\in \mathcal{P}^n}$ and $\{\mathcal{S}^{(j)}(A, \mathcal{RV}_A^\Pi ({\boldsymbol{x}},{\boldsymbol{X}}^{(j)}))\}_{{\boldsymbol{x}}\in (\{0,1\}^{*})^n,{\boldsymbol{X}}^{(j)}\in \mathcal{P}^n}$, $j \in \{1, \hdots, m\}$ satisfy Assumption \ref{ass:comp_en} where $\mathcal{S}^{(j)}$ is defined as in Definition \ref{def:rView}. By Lemma \ref{lemma:comp_en}, 
        $(\mathcal{RV}_A^{\Pi^{(j)}}({\boldsymbol{x}},{\boldsymbol{X}}^{(j)}))_{j=1}^m$ is a refined view for $\Pi$. For statistical and computational refined views, the proofs are similar.
    \end{proof}
    }

}


\bibliographystyle{IEEEtran}
\bibliography{references}

\end{document}